\documentclass[%
 reprint,
 amsmath,amssymb,
 aps,
floatfix,
]{revtex4-2}

\pdfoutput=1

\usepackage{textcomp}
\usepackage{natbib}
\usepackage{graphicx}
\usepackage{siunitx}
\usepackage{xcolor}  

\begin{document}

\title{Revealing the nanogeometry of WS$_2$ nanoflowers by polarization-resolved Raman spectroscopy}

\author{Irina Komen}
\author{Sabrya E. van Heijst}
\author{Martin Caldarola}
\author{Sonia Conesa-Boj}
\author{L. Kuipers\thanks{L.Kuipers@tudelft.nl}}
\date{\today}

\affiliation{Kavli Institute of Nanoscience, Department of Quantum Nanoscience, Delft University of Technology, The Netherlands}

\begin{abstract}
Recent studies of Transition Metal Dichalcogenides (TMDs) have revealed exciting optical properties like stable excitons and chiral light-matter interactions. Chemical vapor deposition (CVD) techniques provide a platform for the fabrication of nanostructures with diverse geometries, ranging from horizontal flakes to flower-like structures. Raman spectroscopy is commonly used to characterize TMDs and their properties. Here, we use polarization-resolved Raman spectroscopy to probe the nanogeometry and orientation of WS$_2$ nanoflower petals. Exciting the nanoflowers with linearly polarized light, we observe an enhanced Raman response from flower petals oriented along the excitation polarization direction. Furthermore, the helicity-resolved Raman response of vertically oriented wall-like flower petals exhibits clear differences with horizontally oriented flakes. Although the photoluminescence from the nanoflowers is strongly reduced, the Raman response upon excitation in resonance with the WS$_2$ excitonic transition does reveal the presence of the exciton, which results in a distinct temperature dependence of the Raman response. 
\end{abstract}

\maketitle

\section{Introduction}

Recently, great scientific interest has been taken in Transition Metal Dichalcogenides (TMDs) and their fascinating optical properties \cite{Mak_MoS2monofirst_PhysRevLett_2010}. TMDs materials like MoS$_2$, WS$_2$, MoSe$_2$ and WSe$_2$, being semiconductors with a bandgap in the visible wavelength range, offer many possibilities for applications in opto-electronics \cite{Zhang_TMDCtransistor_science_2014, Wang_TMDCelectronics_NatNano_2012}. In the TMDs semiconductor valleys, electron and hole pairs form stable excitons even at room temperature \cite{Chernikov_excitonBinding_PRL_2014}. Moreover, the interaction of TMDs with light is chiral, as their pseudospin allows the selective addressing of each TMDs valley by circularly polarized light with opposite handedness \cite{Cao_TMDCcircular_NatCom_2012, Xu_TMDCspins_NatPhys_2014, Zhu_WS2bilayerValleyPolarization_PNAS_2014}. 

Chemical Vapor Deposition (CVD) provides a flexible platform for the fabrication of TMD nanostructures \cite{Song_CVDgrownWS2_ACSNano_2013, Zhang_CVDgrownWS2_ACSNano_2013, Cong_CVDgrownWS2_AdvOptMat_2014, Orofeo_CVDgrownWS2_APL_2014, Thangaraja_WS2crystals_MatLett_2015, Liu_CVDgrownWS2_NanoscResLett_2017}. While CVD can reproduce naturally occurring flat layered TMDs, it also offers the possilibity of fabricating vertical TMDs walls \cite{Jung_verticalTMD_NanoLett_2014}, pyramids \cite{Irina_pyramids_2020} and flower-like nanostructures \cite{Li_MoS2flowers_APL_2003, Li_TMDflowers_chem_2004, Prabakaran_WS2flowers_chemCom_2014, Sabrya2020}. Potential applications of flower-like TMDs structures range from catalysis \cite{Sabrya2020, Prabakaran_WS2flowers_chemCom_2014} to using their large field emission as a potential electron source \cite{Li_MoS2flowers_APL_2003, Li_TMDflowers_chem_2004}. However, so far TMDs nanoflowers have mainly been studied using electron microscopy tools \cite{Sabrya2020}, and little is known about their interaction with light. It is interesting to note that, in contrast to flat layers, no PL but only a Raman response is reported from vertical TMDs walls \cite{Jung_verticalTMD_NanoLett_2014, Fu_verticalWS2polarization_OptLett_2014}, TMDs pyramids \cite{Irina_pyramids_2020} or flower-like TMDs structures \cite{Li_MoS2flowers_APL_2003, Li_TMDflowers_chem_2004, Prabakaran_WS2flowers_chemCom_2014}.

Raman spectroscopy offers a powerful and non-invasive tool for the investigation of TMDs materials \cite{Lee_MoS2Ramanfirst_ACSNano_2010,  Zhao_RamanTMDlinear_Nanoscale_2013, Berkdemir_RamanWS2_ScientRep_2013,  Molas_RamanWS2_ScientRep_2017}. Commonly studied in TMDs are the characteristic vibrational modes, namely the E$^1_{2g}$ that corresponds to the in-plane displacement of the atoms and the A$_{1g}$ that corresponds to the out-of-plane displacement of the chalcogenide atoms, as well as the longitudinal acoustic phonon LA(M). Interestingly, the TMDs' Raman response is highly enhanced when the excitation is on resonance with an excitonic transition \cite{Berkdemir_RamanWS2_ScientRep_2013, McDonnell_resonantRamanWS2_NanoLetters_2018, Corro_resonantRamanTMD_NanoLetters_2016, Golasa_multiphononMoS2_APL_2014}. As this resonance effect can be observed in the Raman response even in the absence of photoluminescence, resonance Raman spectroscopy offers a way to study the TMDs exciton indirectly \cite{Irina_pyramids_2020}. As the TMDs bandgap energy depends on temperature, varying the temperature of a TMD material enables the tuning of the resonance condition for a fixed excitation frequency. Therefore, studying TMDs at various cryogenic temperatures provides insights on the influence of the excitonic transition on the Raman response. Moreover, temperature-dependent Raman spectroscopy can shed light on the structural properties of TMDs materials \cite{Fan_resonanceRamanTMD_JApplPhys_2014, Gaur_temperatureRamanWS2_PhysChemC_2015}. 

The Raman response of TMDs is influenced by the polarization of the excitation light, where the in-plane and the out-of-plane vibrations of the atoms respond differently to either orthogonal, in-plane polarization \cite{Zhao_RamanTMDlinear_Nanoscale_2013}. Furthermore, given the chirality of the TMDs valleys and the resonant influence of the excitons on the Raman response, studying the interaction of TMDs phonons with circularly polarized light is important \cite{Chen_helicityRamanTMD_NanoLett_2015, Zhao_helicityMoS2_ACSNano_2020, Drapcho_helicityTMD_PRB_2017}. As the Raman effect depends on the polarizability of the material, the interaction of TMDs with polarized light is described by a Raman polarizability tensor \cite{Zhao_helicityMoS2_ACSNano_2020, Jin_MoSe2polarization_2020, Ding_RamanTensorsMoS2_optlett_2020}. It is important to note that these tensors are defined with respect to the atomic axes, \textit{e.g.}, typically assuming flat-layered TMDs with the excitation light perpendicular to it. Thus, the polarization-resolved Raman response of for instance a vertical TMDs wall will be completely different than that of a flat layer, \textit{e.g.} modes that are usually allowed/forbidden in cross-polarization will now be absent/observed \cite{Jin_MoSe2polarization_2020, Ding_RamanTensorsMoS2_optlett_2020, Hulman_MoS2polarizationVertical_PhysChemC_2019, Fu_verticalWS2polarization_OptLett_2014}. Therefore, polarization-resolved Raman studies will provide insight in the flowers' nanogeometry and orientation. 

\begin{figure*}[htp]
\centering
\includegraphics[width = 0.65\linewidth] {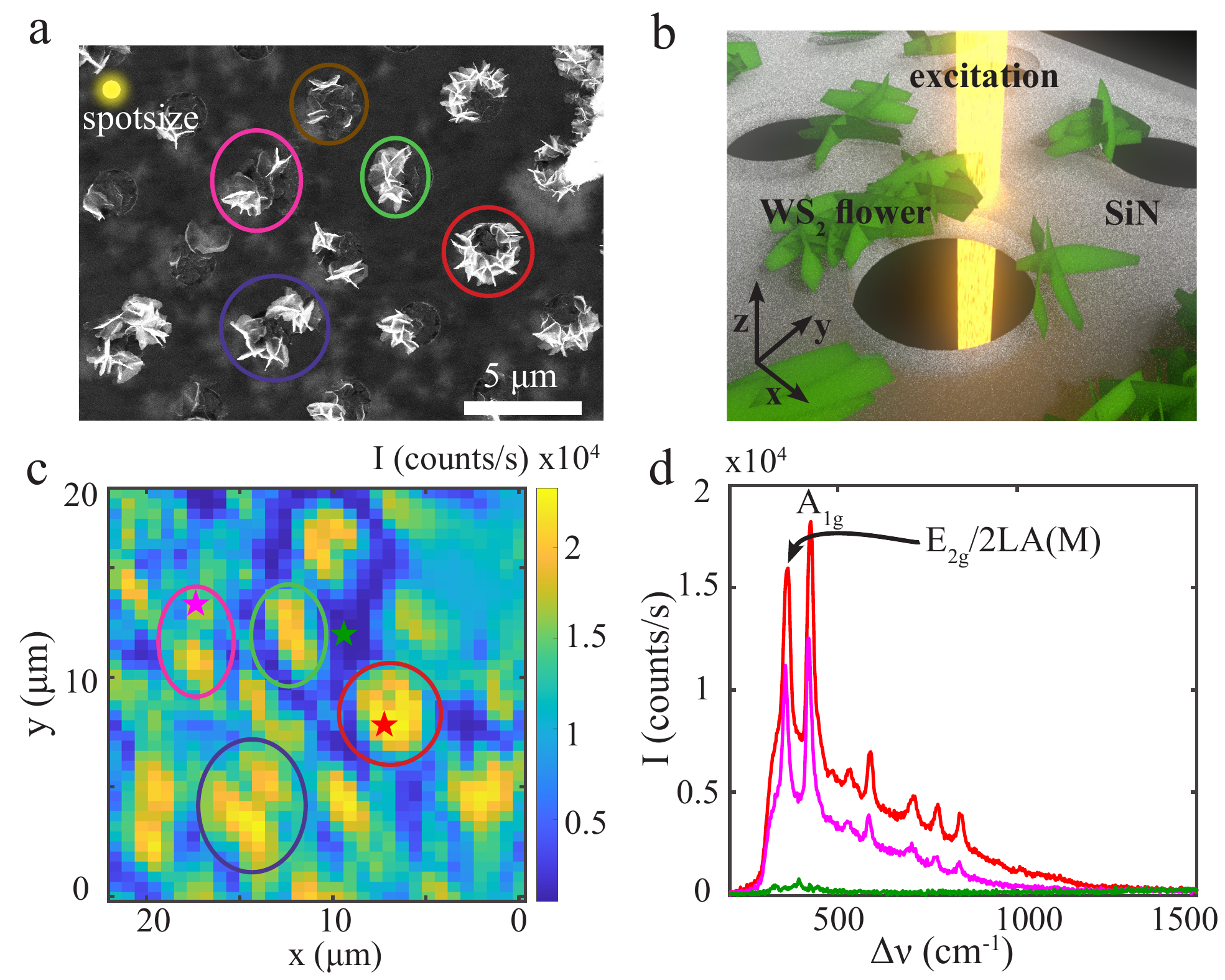}
\caption{\textbf{Optical response of WS$_2$ nanoflowers} \\ \textbf{a.} SEM image of the WS$_2$ nanoflowers on a SiN membrane with circular holes. The flowers grow mainly around the holes, forming diverse flower-like shapes ranging from circles (red) and half-circles (green) to vertical walls (brown, pink) and more chaotic structures (purple, pink). In yellow the size of the excitation laser spot (\SI{500}{\mu m}). \textbf{b.} A schematic representation of the SiN substrate (grey) with holes (black), WS$_2$ nanoflowers (green) and the excitation laser (yellow). \textbf{c.} Map of the peak intensity of the first Raman feature (denoted with an arrow in \textbf{d}), where the shape of the flowers can be clearly correlated with the SEM image in \textbf{a} (see colored circles as guide to the eye). \textbf{d.} Spectra at different positions indicated with stars in \textbf{c} on flowers (red and pink) and on the substrate (green). Note that, even though the flowers have a diversity in shapes, the only difference between spectra of flowers is the intensity the Raman features.}
\label{fig_intro}
\end{figure*}


In this work, we study the polarization- and temperature-dependent optical response of WS$_2$ nanoflowers. The nanoflowers exhibit a highly reduced PL enabling the study of the thereby unobscured Raman response. At first glance, no spectral differences are observed between WS$_2$ flowers of different geometry, except for differences in Raman intensity. However, polarization- and helicity-resolved Raman spectroscopy reveals underlying structural differences between flowers. We find that petals of the flowers oriented vertically exhibit a different response to circularly polarized light than more flat flower petals. Moreover, we find that the relative in-plane orientation of the flower petals with respect to the polarization direction of linearly polarized light, affects the optical response. Surprisingly, the polarization- and helicity-dependent behaviour of the characteristic in-plane and out-of-plane WS$_2$ Raman modes is similar, indicating the similarity of the underlying Raman tensors. Studying the temperature-dependent spectral response of WS$_2$ nanoflowers, we observe the influence of the excitonic resonance on the Raman intensity, helicity and the ratio between the two characteristic WS$_2$ Raman features. 
 

\section{Results and Discussion}

\subsection{Optical response of WS$_2$ nanoflowers}

Figure \ref{fig_intro}a depicts a Scanning Electron Microscopy (SEM) image of the WS$_2$ nanoflowers. The flowers are fabricated using CVD on a Si$_3$N$_4$ membrane (\SI{200}{nm} thickness) with an array of holes (\SI{2}{\mu m} radius and \SI{4}{\mu m} pitch, see Figure \ref{fig_other_peaks}a in the Supplementary Materials). Details about the fabrication and an in-depth study of the electronic and crystallographic properties of these nanoflowers are given by Van Heijst \textit{et al} \cite{Sabrya2020}. Just as natural flowers, these WS$_2$ nanostructures consist of randomly oriented flakes (the petals) expanding from a common point. The WS$_2$ nanoflowers arise mainly around the holes in the substrate (see Fig.\ref{fig_intro}a), forming diverse shapes ranging from circles (red) and half-circles (green) to vertical walls (brown, pink) and more complex structures (purple, pink). The larger structure to the right of Fig.\ref{fig_intro}a is probably a conglomeration of WS$_2$ grown around a dust particle. Figure \ref{fig_intro}b schematically presents the nanoflowers (green) around the holes (black) in the substrate (grey). The excitation light is along the z-axis, and the orientation of the flower petals ranges from completely flat (in x-y plane) to standing up (x-z or y-z plane). The petal thickness is estimated to be between 2 and \SI{30}{nm} \cite{Sabrya2020}. The previously performed scanning transmission electron microscopy (STEM) study reveals that the nanoflowers exhibit a crystallographic polytypism 2H/3R \cite{Sabrya2020} (see Section B.2 of the Supplementary Materials for details). 

We investigate the optical response of the WS$_2$ nanoflowers, which consists mainly of a Raman response. Figure \ref{fig_intro}c presents the intensity of the first Raman feature (see arrow in Fig.\ref{fig_intro}d) of the flowers depicted in Fig.\ref{fig_intro}a. The Raman map can be correlated with the SEM image by comparing the shape and relative position of the flowers (\textit{e.g.}, compare the coloured circles in Fig.\ref{fig_intro}a and Fig.\ref{fig_intro}c). Not surprisingly, the more dense flowers, for instance the circular flower (red) and the half-circle (green), exhibit a larger Raman intensity than the structures with mainly upstanding walls (brown, upper part of pink). It is important to note in this context that the size of our diffraction limited excitation spot (\SI{500}{nm}, see Methods) is much larger than the size of an individual flower petal. For an easy comparison, the size of the excitation spot is indicated on scale in yellow in Fig.\ref{fig_intro}a. The Raman signal of the flowers in the Raman map is `smeared out', and the area of plain substrate is actually much larger than it seems on the Raman maps (compare Fig.\ref{fig_intro}c with the SEM image in Fig.\ref{fig_intro}a). The reason for the `smearing out' is that we measure a convolution of the excitation and detection volume with the spatial distribution of the optical response of the flowers. It is to be expected that the spatial distribution of the optical response of the nanoflower response is related to the size of the flower features. 

Figure \ref{fig_intro}d presents optical spectra of the WS$_2$ nanoflowers. The spectra contain of 8-10 Raman features, where the first two features are the characteristic vibrational modes of WS$_2$ (see Figure \ref{fig_determine} in the Supplementary Materials). The first feature is a combination of the in-plane vibrational mode E$_{2g}$ and the longitudinal acoustic phonon 2LA(M) (in WS$_2$, the frequency of these modes is almost the same), and the second feature is the out-of-plane vibrational mode A$_{1g}$. We attribute the higher frequency Raman features to multiphonon resonances involving the LA(M) phonon, excited because the \SI{595}{nm} laser is in resonance with the A-exciton, in accordance with the attribution for WS$_2$ pyramids \cite{Irina_pyramids_2020} (see Section B.1 of the Supplementary Materials for details).  

The spectra in Fig.\ref{fig_intro}d are collected from different positions of the sample: on the Si$_3$N$_4$ substrate (green), on a dense nanoflower (red) and on a vertical-wall nanoflower (pink) (indicated  with stars in Fig.\ref{fig_intro}c). It is interesting to note that the only difference between the red spectrum of the more dense flower and the pink spectrum of the vertical-wall flower is in the overall Raman intensity and not in the spectral position of the Raman peaks. In other words: there are no specific Raman features more or less pronounced for flowers with different nanogeometries.
 

The WS$_2$ nanoflowers exhibit a strongly reduced photoluminescence (PL) with respect to horizontally layered WS$_2$. On some flowers, no PL can be observed from the nanoflowers within our detection efficiency. Specific parts of some nanoflowers do exhibit a low PL, which becomes apparent especially at cryogenic temperatures (see Figure \ref{fig_backgroundPL}d in the Supplementary Materials). At \SI{4}{K}, this is at most \SI{2}{\%} of the PL of a monolayer WS$_2$. Assuming that the absorption and the effective collection efficiency remain constant, we conclude that the CVD grown WS$_2$ nanoflowers have a lower quantum efficiency than horizontal WS$_2$ flakes. Here the assumption of a constant absorption is reasonable given the petal thickness, whereas the assumption of a constant effective collection efficiency is related to the unknown emission pattern from the nanoflower petals and therefore less strong. We attribute the decrease in the quantum efficiency to the increase in possible non-radiative loss channels due to the presence of all the edges of the nanoflower petals. This leads to a severe quenching of the exciton photoluminescence, without influencing the Raman response. 

\begin{figure*}[htp]
\centering
\includegraphics[width = \linewidth] {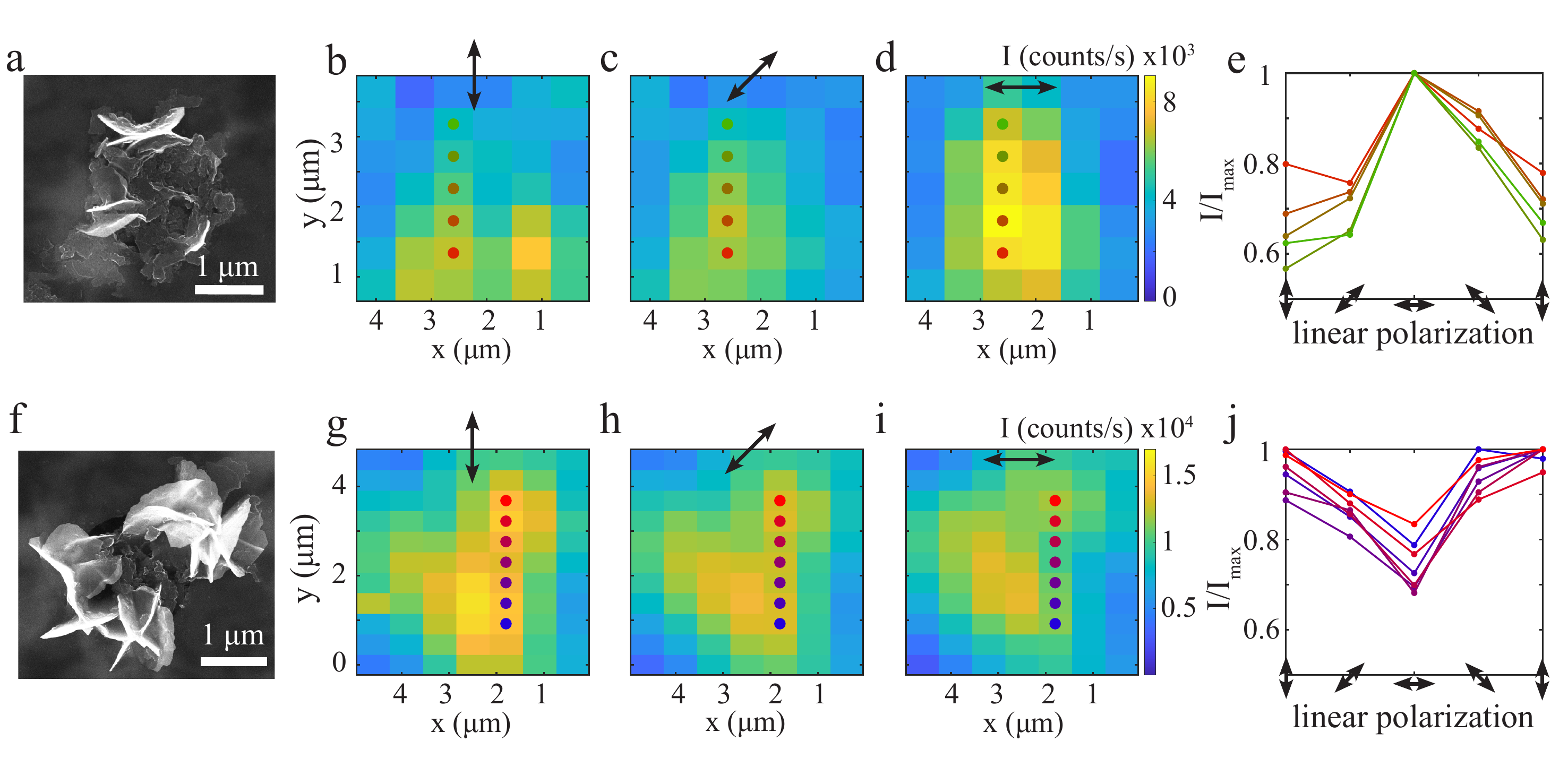}
\caption{\textbf{Excitation polarization} \\ \textbf{a.} SEM image of a WS$_2$ flower-like structure (brown circle in Fig.\ref{fig_intro}a) with mainly petals oriented in the x-z plane. \textbf{b-d,g-h} Map of the intensity of the first Raman feature of the flower-like structure upon \textbf{b,g.} vertically polarized, \textbf{c,h.} diagonally polarized and \textbf{d,i.} horizontally polarized excitation. \textbf{e.} Raman intensity of the flower in \textbf{a} (used pixels are marked with stars in \textbf{b-d}) as a function of excitation polarization angle. Note that the intensity increases drastically when the polarization direction is parallel to the WS$_2$ flower petals. \textbf{f.} SEM image of a WS$_2$ flower (purple circle in Fig.\ref{fig_intro}a) with mainly petals oriented in the y-z plane. \textbf{j.} Raman intensity of the flower in \textbf{f} (used pixels are marked with stars in \textbf{g-i}) as a function of excitation polarization angle. Note that the intensity decreases drastically when the polarization direction is perpendicular to the WS$_2$ flower petals.}
\label{fig_linearPol}
\end{figure*}

\subsection{Polarization-resolved Raman response} \label{sect_flowers_linear_polarization}

\begin{figure*}[hbp]
\centering
\includegraphics[width = 0.9\linewidth] {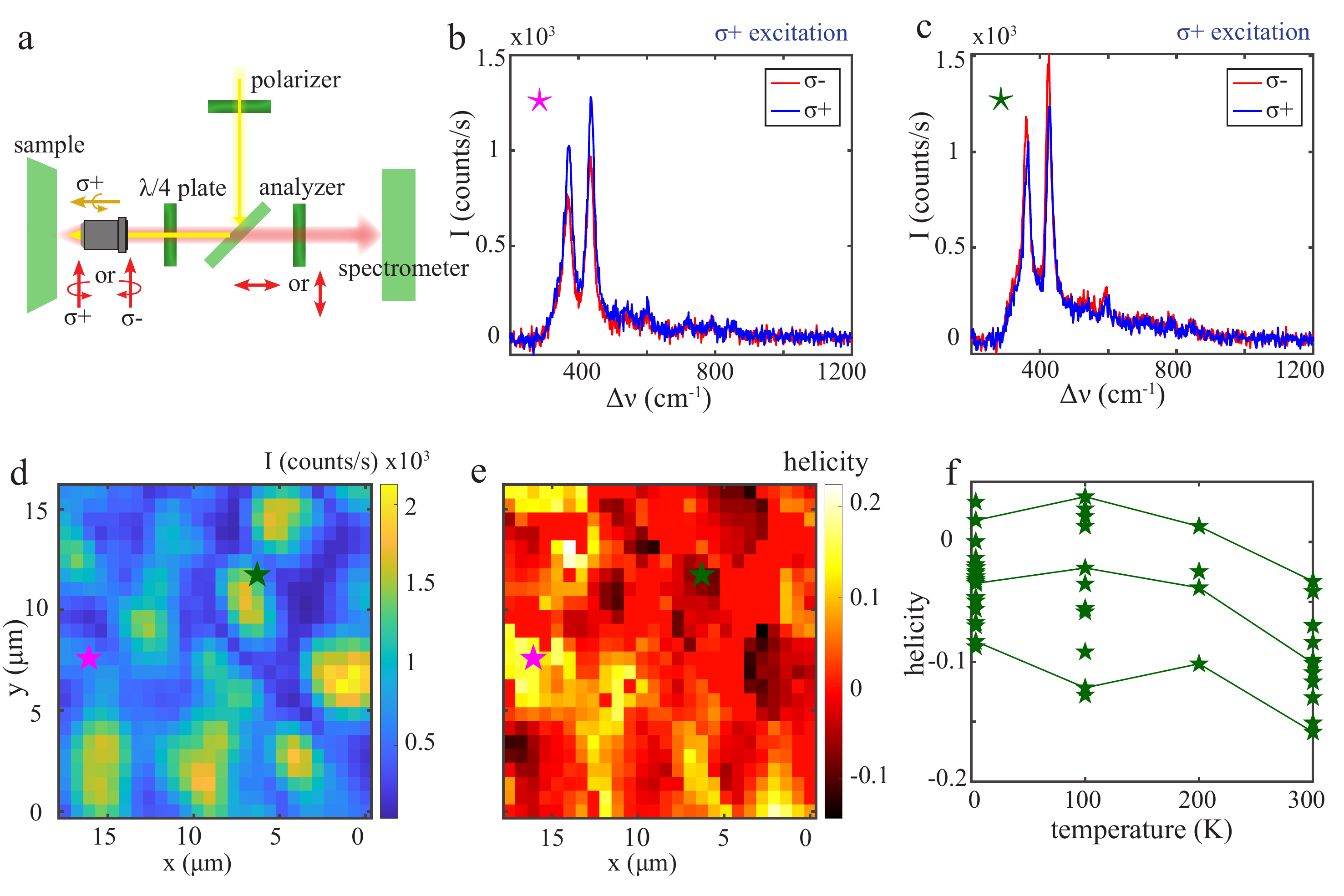}
\caption{\textbf{Helicity of Raman features} \\ \textbf{a.} Schematic of our set-up, where the excitation light (\SI{595}{nm} wavelength) passes through a quarter-wave plate and is focused on the sample. The emission is collected in epi-configuration and passes through the same quarter-wave plate. Than it is directed to a spectrometer through a polarization analyzer. \textbf{b,c} Helicity-resolved nanoflower spectra, where the flowers are excited with $\sigma_+$ light and the helicity is determined from the difference in $\sigma_+$ and $\sigma_-$ emission. In \textbf{b}, the spectrum with the same polarization as the excitation light (blue) has a higher intensity (helicity is conserved). In \textbf{c}, the spectrum with opposite polarization to the excitation light (red) has a higher intensity (helicity is reversed). \textbf{d.} Map of the intensity of the first Raman feature of the nanoflower spectra. \textbf{e.} Map of the same region of the helicity of the first Raman feature. Note that the helicity of the Raman features around the WS$_2$ nanoflowers is negative (green star), whereas the Raman helicity is positive in regions next to the larger nanoflowers (pink star). \textbf{f.} Temperature-dependent helicity of the WS$_2$ nanoflower marked in green in Fig.\ref{fig_intro}a (taking into account all pixels associated with this flower). The lines present the temperature dependence of three locations on the flower marked in green (see Figure \ref{fig_helicity_other}a,b of the Supplementary Materials for the taken pixels). The helicity decreases slightly at room-temperature.}
\label{fig_helicity}
\end{figure*}

To investigate the optical differences between different flowers in more detail, we study the interaction of the flower Raman response with linearly polarized light. Here we excite the WS$_2$ nanoflowers with linearly polarized light, rotating the polarization direction from vertical to horizontal, and analyze the resulting emission intensity (see Fig.\ref{fig_helicity}a for a schematic of our set-up, where the quarter-wave plate and polarization analyzer are not used in the current section). 

Figure \ref{fig_linearPol}a depicts an SEM image of a flower-like WS$_2$ structure (indicated in brown in Fig.\ref{fig_intro}a) with mainly wall-like petals, oriented in the x-z plane (see coordinate system in Fig.\ref{fig_intro}b). Figures \ref{fig_linearPol}b-d depict the intensity of the first Raman feature upon vertical polarization excitation, excitation polarization at \SI{45}{degrees} and horizontal polarization excitation. The Raman intensity is highest when the excitation polarization direction is parallel to the orientation of the nanoflower petals, in this case upon horizontal excitation (Fig.\ref{fig_linearPol}d). This becomes even more apparent in Fig.\ref{fig_linearPol}e, where the normalized Raman intensity of different parts of the nanoflower (positions are indicated in Fig.\ref{fig_linearPol}b-d) is plotted as a function of polarization angle (depicted by the arrows). The Raman intensity upon vertical polarization is 60 - 80 \% of the Raman intensity upon horizontal polarization. Note in Fig.\ref{fig_linearPol}b that the small flower petal to the right of the flower, oriented vertically in the y-z plane, can only be distinguished upon vertical polarization: it is not visible anymore in Fig.\ref{fig_linearPol}c and d. 

To illustrate the correlation between the Raman intensity of differently oriented flower-like structures and the excitation polarization even more, Fig.\ref{fig_linearPol}f depicts a nanoflower (indicated in purple in Fig.\ref{fig_intro}a) which exhibits petals oriented in the y-z plane (see the coordinate system in Fig.\ref{fig_intro}b). Here, the Raman intensity upon vertical y polarization excitation (Fig.\ref{fig_linearPol}g) is higher than upon horizontal x polarization excitation (Fig.\ref{fig_linearPol}i). Figure \ref{fig_linearPol}j depicts the normalized Raman intensity of different parts of the nanoflower (positions are indicated in Fig.\ref{fig_linearPol}g-i) as a function of polarization angle (depicted by the arrows). For this flower, the Raman intensity upon horizontal excitation is now \mbox{70 - 90 \%} of the Raman intensity upon horizontal polarization. The lower contrast can be explained by the fact that this nanoflower is more dense, also containing petals oriented differently than strictly in the y-z plane, which demonstrates the sensitivity of this method. Flowers with petals oriented in random different directions do not exhibit a polarization dependence (see Figure \ref{fig_polarization_independence} in the Supplementary Materials). 

The response of Raman modes to polarized light is described by Raman polarizability tensors, based on the crystal symmetries in the material \cite{Ding_RamanTensorsMoS2_optlett_2020, Hulman_MoS2polarizationVertical_PhysChemC_2019, Fu_verticalWS2polarization_OptLett_2014, Jin_MoSe2polarization_2020}. It is interesting to point out that the measured E$_{2g}$ and A$_{1g}$ Raman features exhibit the same polarization response (see Figure \ref{fig_polarization_modes} in the Supplementary Materials). This indicates that the Raman polarization tensor for both the in-plane (E$_{2g}$) and the out-of-plane (A$_{1g}$) Raman modes are the same. We also found that the polarization dependence of the Raman intensity does not depend on temperature and is also observed upon \SI{561}{nm} excitation (see Figure \ref{fig_polarization_modes} in the Supplementary Materials). We conclude that linear-polarization-resolved Raman measurements provide a way to distinguish between differently oriented WS$_2$ petals and to identify the dominant orientation. 




\subsection{Helicity of Raman features}

Another tool to investigate potential optical differences between nanoflowers with diverse geometries, is helicity-resolved Raman measurements. \mbox{Figure \ref{fig_helicity}a} depicts a schematic representation of our set-up. The excitation light (\SI{595}{nm} wavelength) passes through a quarter-wave plate and is focused on the sample through an objective lens (see Methods for details on the set-up). The emission is collected through the same objective lens, passes a quarter-wave plate, and is directed to a spectrometer through a polarization analyzer. This allows the detection of the polarization state of the emitted light, i.e., it allows for helicity-resolved measurements. \mbox{Figures \ref{fig_helicity}b,c} depict helicity-resolved nanoflower spectra. Here, the flowers are excited with $\sigma_+$ circularly polarized light and the helicity of the Raman features is determined from the difference in $\sigma_+$ and $\sigma_-$ emission. In Fig.\ref{fig_helicity}b, the blue spectrum with the same polarization as the excitation light, has a higher intensity ($\sigma_+$, helicity is conserved) than the red spectrum with the opposite polarization ($\sigma_-$, helicity is reversed). We calculate the helicity of the first Raman feature $H = \frac{I_{conserved} - I_{reversed}}{I_{conserved} + I_{reversed}}$ to be 0.172. In Fig.\ref{fig_helicity}c, the helicity-reversed spectrum (red) has a higher intensity than the helicity-conserved spectrum (blue), with H = -0.083. 
The helicity of the Raman response of the WS$_2$ nanoflowers is position dependent. Figure \ref{fig_helicity}d presents a map of the nanoflower intensity of the first Raman feature (compare Fig.\ref{fig_intro}c). Figure \ref{fig_helicity}e presents a map of the experimentally determined helicity of the first Raman feature (stars indicate the position of spectra in Fig.\ref{fig_helicity}b,c). Note again that the measured position-dependent Raman intensity and helicity are a convolution of the excitation and detection volume with the spatial distribution of the optical response of the flowers, related to the size of the flower features. The Raman helicity of the WS$_2$ nanoflowers is negative: the intensity is higher for the helicity-reversed spectrum. Note however that the locations where the most negative Raman helicity is located, is not in the middle of the nanoflower, but towards the edge (\textit{e.g.}, compare the green star in Fig.\ref{fig_helicity}e and Fig.\ref{fig_helicity}d). We therefore conclude that we detect a negative Raman helicity at locations where the excitation spot interacts with the side of a nanoflower. The helicity is most positive on locations in between the WS$_2$ nanoflowers, for instance at the position of the pink star: here the intensity is higher for the helicity-conserved spectrum. The Raman response from these regions confirms the presence of WS$_2$, e.g., this is not the bare substrate. Comparing the position of the pink star in Fig.\ref{fig_helicity}d with the SEM image in Fig.\ref{fig_intro}a, it seems that the region of positive helicity is actually related to the WS$_2$ structure to the left of the flower indicated in purple in Fig.\ref{fig_intro}a. As this structure looks more flat than the wall-like petals in other flowers, we conclude that the sign of the Raman helicity becomes positive when the WS$_2$ is oriented in the x-y plane, horizontally with respect to the surface (see Fig.\ref{fig_intro}b for a coordinate system). 

The Raman helicity response of the WS$_2$ nanoflowers is completely different than that of flat layers of WS$_2$. As alluded to before, the response of Raman modes to polarized light is described by Raman tensors \cite{Zhao_helicityMoS2_ACSNano_2020, Jin_MoSe2polarization_2020, Ding_RamanTensorsMoS2_optlett_2020} (see Section E of the Supplementary Materials). In case of TMDs materials, the Raman tensor dictates that the A$_{1g}$ mode is helicity-conserved \cite{Zhao_helicityMoS2_ACSNano_2020, Chen_helicityRamanTMD_NanoLett_2015}. This means that the second Raman feature in \ref{fig_helicity}b,c should only have had contributions with the same polarization as the excitation ($\sigma_+$), leading to H = 1.0. However, we observe a large contribution of light with the reversed helicity, in Fig.\ref{fig_helicity}c the helicity even becomes negative in places (see Figure \ref{fig_helicity_other} in the Supplementary Materials for a helicity map of the A$_{1g}$ mode). 
Interpreting the helicity behaviour of the first Raman feature is less straightforward, as this feature contains both the 2LA(M) phonon and the E$_{2g}$, and the Raman tensor of the E$_{2g}$ depends on the resonance of the excitation. The tensor dictates that the E$_{2g}$ mode is helicity-reversed under non-resonant excitation \cite{Chen_helicityRamanTMD_NanoLett_2015} and helicity-conserved under resonant excitation \cite{Zhao_helicityMoS2_ACSNano_2020, Drapcho_helicityTMD_PRB_2017} (see Section E.2 of the Supplementary Materials). Since the nanoflowers are excited at resonance with the excitonic energy, the first Raman feature in \ref{fig_helicity}b,c should have had mainly contributions with the same polarization as the excitation. Therefore the resonance of the excitation explains why the E$_{2g}$ and the A$_{1g}$ features have a similar helicity \cite{Zhao_helicityMoS2_ACSNano_2020}. However, the observation of negative helicity is surprising for both Raman features, as the response is completely different than that of flat WS$_2$ layers. 

It is important to note that the Raman polarization tensors are typically defined with respect to the crystal axes of flat TMDs layers, which for flat layers are readily connected to a suitable frame of reference of the incident light. The petals of the WS$_2$ nanoflowers exhibit a variety of orientations with respect to the incident light. Mathematically, a change of WS$_2$ flake orientation corresponds to a base transformation changing the Raman tensor, which may lead to allowed modes becoming forbidden and forbidden modes becoming allowed (see Figure \ref{fig_base_transformation} of the Supplementary Materials). From Fig.\ref{fig_helicity}e it is apparent that the Raman helicity of the WS$_2$ nanoflowers is in general slightly negative, with a larger helicity-reversed than helicity-conserved contribution. This corresponds to the nanoflowers on average having more wall-like petals (oriented in x-z or y-z plane, see Fig.\ref{fig_intro}b for a coordinate system), which is in agreement with the SEM images of the flowers. However, the fact that the helicity is at most -0.2 indicates that the contribution of both flat and vertically oriented flower petals within the diffraction-limited excitation spot is relatively large. 

\begin{figure*}[htp]
\centering
\includegraphics[width = 0.65\linewidth] {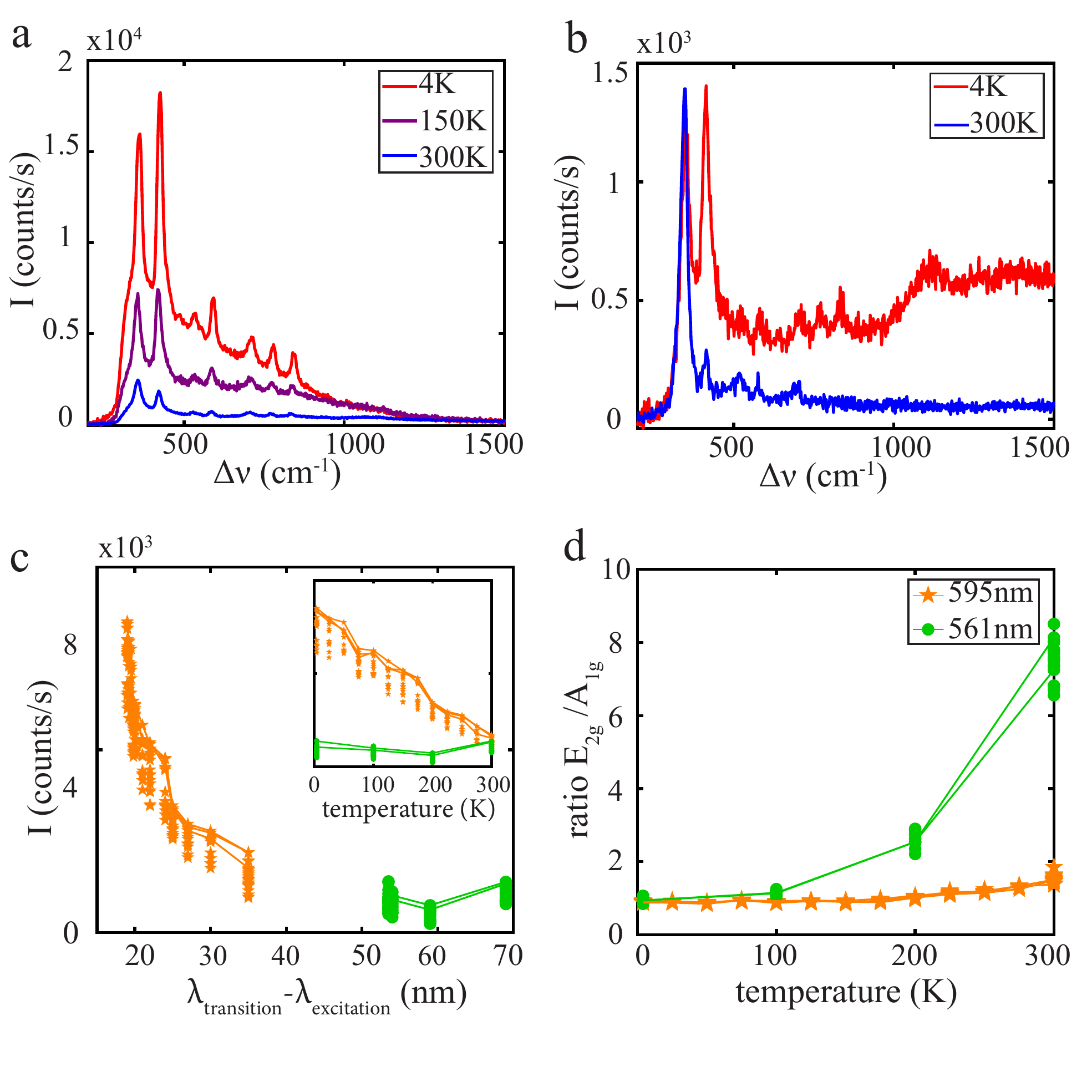}
\caption{\textbf{Temperature dependence Raman intensity} \\ \textbf{a.} Nanoflower spectra (flower indicated in red in Fig.\ref{fig_intro}a) upon a \SI{595}{nm} excitation at temperatures ranging from \SI{4}{K} to room-temperature. Note that the intensity of all Raman features increases. \textbf{b.} Nanoflower spectra (flower indicated in red in Fig.\ref{fig_intro}a) upon a \SI{561}{nm} excitation at room-temperature and \SI{4}{K}. Note that the A$_{1g}$ mode is almost absent at room temperature. \textbf{c.} \textit{inset} Temperature-dependent intensity of the first Raman feature (E$_{2g}$,2LA(M)) of the nanoflower spectra upon \SI{595}{nm} excitation (orange) and \SI{561}{nm} excitation (green). Here at every temperature the intensity is taken from all pixels associated with the flower (indicated in red in Fig.\ref{fig_intro}d). \textit{main} The Raman intensity is plotted as a function of the wavelength difference between the WS$_2$ bandgap and the excitation. Upon cooling down, the WS$_2$ bandgap energy blue shifts. With a constant excitation energy, the difference between excitation and WS$_2$ bandgap energy will become smaller at lower temperatures, bringing the excitation more in resonance with the excitonic transition. \textbf{d.} Temperature-dependent ratio of the first two Raman features of the nanoflower spectra. Upon a \SI{595}{nm} excitation, the ratio changes from 0.8 at 4K to 1.6 at room-temperature, as can already seen by comparing the intensity of the first two Raman features in \textbf{a}. Upon a \SI{561}{nm} excitation, the A$_{1g}$ is almost absent at room-temperature. Therefore, the ratio between the two WS$_2$ flower Raman features increases drastically from 1.0 at 4K to 7.5 at room-temperature. }
\label{fig_temperature}
\end{figure*}

Based on the Raman tensor, flat flower petals (oriented in the x-y plane) should exhibit a positive helicity (see Section E of the Supplementary Materials). Comparing the helicity map with the SEM image in Fig.\ref{fig_intro}a, it is not always straightforward to correlate the regions of positive helicity with the orientation and nanogeometry of flower petals. We hypothesise that there might be flat flakes present that cannot be clearly distinguished from the Si$_3$N$_4$ substrate, but that do contribute to the positive Raman helicity. We conclude that the surprising helicity values for the nanoflower Raman response can be explained by the different orientations of the flower petals. 

We determine the position-dependent helicity of the Raman features at different temperatures (see Figure \ref{fig_helicity_300K} in the Supplementary Materials). Figure \ref{fig_helicity}f depicts the temperature dependence of the Raman helicity of the flower marked in green in  Fig.\ref{fig_intro}a,c. At all temperatures, the intensity is depicted of the first Raman feature of all spectra associated to this flower. The lines present the temperature dependence of three specific places on the flower marked in green (see Figure \ref{fig_helicity_other}a,b in the Supplementary Materials). The helicity at room temperature seems to be slightly lower than the helicity at cryogenic temperatures, but the trend is not clear. The helicity of the A$_{1g}$ mode and of the first Raman feature of spectra of other flowers also decreases at room temperature (see Figure \ref{fig_helicity_other}d,e in the Supplementary Materials). The lower helicity at room temperature can be explained by the excitation energy being more out-of-resonance with the excitonic bandgap energy (see Fig.\ref{fig_temperature}). We conclude that the main mechanism that determines the Raman helicity is the flower petal orientation and therefore independent of temperature. Therefore helicity-dependent Raman spectroscopy can be used to determine the orientation of WS$_2$ flakes and the contribution of flat vs. wall-like petals in WS$_2$ nanoflowers.

\subsection{Temperature-dependent Raman spectroscopy}

Given the phononic nature of Raman scattering, studying the temperature dependence of the Raman spectra of the WS$_2$ nanoflowers provides valuable information. \mbox{Figures \ref{fig_temperature}a,b} present the spectral response of the flower indicated in red in Fig.\ref{fig_intro}a (see Figure \ref{fig_polarization_independence} of the Supplementary Materials for an SEM image), upon a \SI{595}{nm} and a \SI{561}{nm} excitation at different temperatures. There are 8-10 Raman features distinguishable at room temperature and at cryogenic temperatures (see Figure \ref{fig_determine} in the Supplementary Materials), but the intensity of the features increases drastically with decreasing temperature. At \SI{4}{K} there is a broad background visible under the Raman features (at 200 - \SI{700}{cm^{-1}} in Fig.\ref{fig_temperature}a and at 1200 - \SI{1500}{cm^{-2}} in Fig.\ref{fig_temperature}b). We attribute this background to highly reduced WS$_2$ photoluminescence (see Section B.1 of the Supplementary Materials). The intensity of the Raman features is much lower for the \SI{561}{nm} excitation than for the \SI{595}{nm} excitation. This is attributed to the fact that the \SI{595}{nm} excitation light is close to the A-exciton resonance of WS$_2$, whereas the \SI{561}{nm} is out-of-resonance with the A-exciton. Raman modes of TMDs can be greatly enhanced when they are excited in resonance with an excitonic transition \cite{Berkdemir_RamanWS2_ScientRep_2013,Zhao_RamanTMDlinear_Nanoscale_2013,  Corro_resonantRamanTMD_NanoLetters_2016, McDonnell_resonantRamanWS2_NanoLetters_2018}. 

The inset of Fig.\ref{fig_temperature}c depicts the temperature-dependent intensity of the first Raman feature (E$_{2g}$,2LA(M)) upon \SI{595}{nm} excitation (orange) and \SI{561}{nm} excitation (green). Here, for every temperature the Raman intensity of all the spectra associated to the nanoflower are taken (flower indicated in red in Fig.\ref{fig_intro}c). The lines present the temperature dependence of three specific places on the flower. For an excitation at \SI{595}{nm}, the Raman intensity decreases with increasing temperature, but for an excitation at \SI{561}{nm}, the Raman intensity is independent of temperature. Figure \ref{fig_temperature}c depicts the intensity of the first Raman feature as a function of the difference between the WS$_2$ exciton and the excitation wavelength. The WS$_2$ bandgap energy and therefore the exciton energy is temperature dependent, experiencing a blue shift with decreasing temperature (see Figure \ref{fig_backgroundPL} of the Supplementary Materials). Therefore varying the temperature of the WS$_2$ nanoflowers enables the tuning of the exciton resonance condition for a fixed excitation frequency. Since the exciton energy is experiencing a blue shift with decreasing temperatures, cooling down the WS$_2$ nanostructures will bring the excitation more in resonance with the excitonic transition. It is clear in Fig.\ref{fig_temperature}c that the Raman intensity exhibits a resonant-like enhancement as the excitation wavelength approaches the excitonic transition. Since the \SI{561}{nm} excitation is relatively far away from the WS$_2$ bandgap, the resonance effect on the Raman intensity upon cool down is much less visible. 



When comparing the spectra upon a \SI{595}{nm} excitation in Fig.\ref{fig_temperature}a, it becomes apparent that the ratio between the two characteristic WS$_2$ Raman features ($E_{2g}/A_{1g}$), is temperature dependent. At room-temperature, the $E_{2g}$ mode is 1.5 times as intense as the $A_{1g}$ mode, and at \SI{4}{K}, the $A_{1g}$ mode is 1.5 times as intense as the $E_{2g}$ mode. It has been reported before, that the different TMDs Raman modes respond differently to the excitonic resonance \cite{Carvalho2015, McDonnell_resonantRamanWS2_NanoLetters_2018, Corro_resonantRamanTMD_NanoLetters_2016}. When comparing the nanoflower spectra upon \SI{561}{nm} excitation in Fig.\ref{fig_temperature}b, the low intensity of the second Raman feature ($A_{1g}$) at room-temperature draws immediate attention. Figure \ref{fig_temperature}d depicts the temperature dependence of the $E_{2g}/A_{1g}$ ratio. At room temperature, the ratio between the characteristic WS$_2$ Raman features is around 7.0, for an excitation at \SI{561}{nm}. From Fig.\ref{fig_temperature}d we deduce that the $A_{1g}$ Raman feature is more sensitive to the resonance conditions than the $E_{2g}$,2LA(M) feature. Even if the \SI{561}{nm} excitation is relatively far away from the exciton wavelength, the A$_{1g}$ Raman mode is enhanced greatly at cryogenic temperatures, as the excitation is closer to the excitonic resonance. Therefore we conclude that the absence of photoluminescence does not prevent an indirect study of the exciton, the presence of which is revealed by resonant Raman spectroscopy. 

\section{Conclusion}

We have studied the optical response of CVD grown WS$_2$ nanoflowers. In contrast to flat WS$_2$ flakes, the nanoflowers exhibit a highly reduced photoluminescence enabling the study of their clear Raman response. Even though the WS$_2$ exciton emission is reduced in the nanoflowers, the presence of the excitons is still notable in the Raman response upon resonance excitation. We study the temperature-dependent Raman intensity and observe an enhancement for cryogenic temperatures, where the intensity of the out-of-plane Raman mode A$_{1g}$ is enhanced more than the intensity of the in-plane Raman mode E$_{2g}$. We conclude that, due to the temperature-dependent bandgap and thus exciton energy shift, the WS$_2$ nanoflowers are excited more in resonance with the excitonic transition at cryogenic temperatures, leading to a resonant effect on the Raman intensity. 

Furthermore, we study the interplay between flower geometry and spectral response. Even though the WS$_2$ nanoflowers have completely different geometries, at first sight the only spectral differences between them seem to be the Raman intensity. However, helicity-resolved and polarization-resolved Raman spectroscopy reveals underlying structural and geometrical differences between flowers. Studying the Raman response upon excitation with circularly polarized light reveals a completely different behaviour of the Raman helicity of the flowers with respect to flat WS$_2$ flakes. The Raman helicity  of nanoflowers with many vertical walls is slightly negative, and the Raman response of flat lying WS$_2$ flower petals is slightly positive. We attribute the differences between the nanoflowers and the flat WS$_2$ to a difference in the Raman polarization tensor, induced by the differently oriented flower petals. Studying the Raman response upon excitation with linearly polarized light we observe that we can selectively address nanoflower petals oriented parallel to the used polarization. We conclude that there is a interplay between the orientation of the flower petals, the atomic vibrational modes and the polarization direction of the excitation light. 

Therefore we envision that temperature-dependent Raman spectroscopy will open the way to study excitonic resonance effects, and polarization-resolved Raman spectroscopy will open the way to determine the nanogeometry and orientation of WS$_2$ flakes. 

\section{Experimental Section}

The WS$_2$ nanoflowers are directly grown on a microchip using chemical vapour deposition (CVD) techniques. The sample preparation method is described in \cite{Sabrya2020}. 
The optical measurements are performed using a home-built spectroscopy set-up, depicted schematically in Fig.\ref{fig_helicity}a. The sample is placed on a piezo stage in a Montana cryostation S100. Measurements are performed at a range of temperatures between room temperature and \SI{4}{K}. The sample is illuminated through an \SI{0.85}{NA} Zeiss 100x objective. Measurements are performed using a continuous wave laser with a wavelength of \SI{595}{nm} and a power of \\ \SI{1.6}{mW/mm^2} (Coherent OBIS LS 594-60), and the excitation light is filtered out using colour filters (Semrock NF03-594E-25 and FF01-593/LP-25). For the measurements depicted in Fig.\ref{fig_temperature}, a continuous wave laser with a wavelength of \SI{561}{nm} and a power of \SI{3.6}{mW/mm^2} is used (Cobolt 08-01/561). To avoid the depolarization consequences of tight focusing on (circular) polarization, a \SI{2}{mm} laser diameter is used, slightly underfilling the objective in the excitation path. Polarizers (Thorlabs LPVIS100-MP2) and superachromatic waveplates are used to rotate the linear polarization (Thorlabs SAHWP05M-700) and create circular polarization (Thorlabs SAQWP05M-700), respectively. The sample emission is collected in reflection through the same objective as in excitation, and projected onto a CCD camera (Princeton Instruments ProEM 1024BX3) and spectrometer (Princeton Instruments SP2358) via a 4f lens system. 

\section{Acknowledgements} 
M.C. acknowledges the financial support of the Kavli Institute of Nanoscience Delft through the KIND fellowships program. S.C.B and S.v.H. acknowledge funding from ERC Starting Grant “TESLA” No. 805021.

\clearpage

\section*{Supplementary Materials}

\subsection{Remnant photoluminescence}

\begin{figure*}[htp]
\centering
\includegraphics[width = 0.9\linewidth] {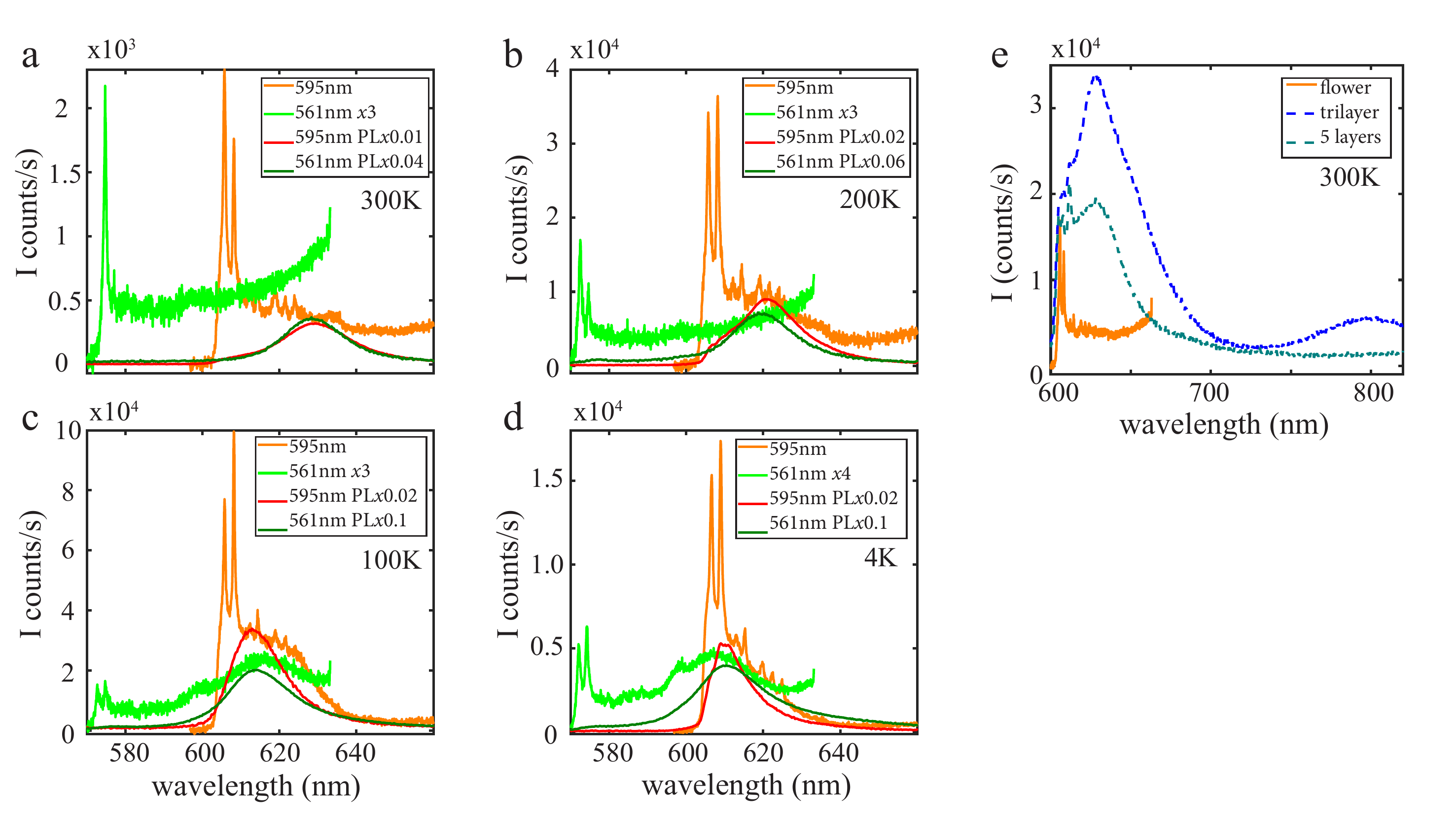}
\caption{\textbf{Comparison spectral response flower and monolayer WS$_2$} \\
\textbf{a-d.} Temperature-dependent spectral response of a WS$_2$ nanoflower upon \SI{595}{nm} excitation (orange) and \SI{561}{nm} excitation (light green), compared with the photoluminescence (PL) response of a monolayer WS$_2$ upon \SI{595}{nm} excitation (in red) and \SI{561}{nm} excitation (dark green), rescaled for an easy comparison (see legends). The spectral response of the nanoflowers contains both sharp Raman features and a broad background. In \textbf{c-d}, this broad background around \SI{615}{nm} overlaps spectrally with the sharp Raman features upon \SI{595}{nm} excitation (in orange), but is well separated from the Raman features upon \SI{561}{nm} excitation (light green). Especially in the spectra at \SI{4}{K} and \SI{100}{K} (\textbf{c-d}), the spectral background under the Raman features of the nanoflowers is at the same spectral position as the monolayer PL. At all temperatures, the remnant PL in the nanoflower spectra is at most 2\% of the monolayer PL. \textbf{e.} (Room-temperature) spectra of a WS$_2$ trilayer (in blue) and five layers of WS$_2$ (in blue-green) (exfoliated on a Si substrate), compared with a spectrum of a WS$_2$ nanoflower (in orange). For few-layer WS$_2$, the PL intensity from the direct transition is reduced by an order of magnitude with respect to a monolayer, but is still clearly distinguishable from the background. The intensity of the remnant PL from the nanoflower is however another order of magnitude lower than the PL of few-layer WS$_2$.}
\label{fig_backgroundPL}
\end{figure*}

As mentioned in the main text, the spectral response of the WS$_2$ nanoflowers exhibits 8-10 Raman features, in combination with a broad background. Figure \ref{fig_backgroundPL}a-d presents temperature-dependent spectra of a nanoflower upon \SI{595}{nm} excitation (in orange) and \SI{561}{nm} excitation (in light green). From Fig.\ref{fig_backgroundPL}c,d it becomes apparent, that for the spectra upon \SI{595}{nm} excitation, the maximum of the broad background is found at approximately \SI{615}{nm} and overlaps spectrally with the sharp Raman features. For the spectra upon \SI{561}{nm} excitation, the broad background is well separated from the Raman features. The spectral position of the broad background is the same for both excitations. 

As a comparison, Fig.\ref{fig_backgroundPL}a-d also present temperature-dependent PL spectra of a WS$_2$ monolayer upon \SI{595}{nm} excitation (in red) and \SI{561}{nm} excitation (dark green). The spectral position of the PL shifts from \SI{630}{nm} at room temperature to \SI{615}{nm} at cryogenic temperatures. At \SI{100}{K} and \SI{4}{K} it is clearly visible, that the broad background under the Raman features of the nanoflowers is at the same spectral position as the monolayer PL spectra. The maximum peak intensity of the broad background is around 2\% of the PL intensity. At room temperature and \SI{200}{K}, the background under the Raman features is broader than the monolayer PL spectra. At these temperatures, the intensity of the background is not higher than 2\% of the monolayer photoluminescence. As the depicted spectra are taken from nanoflowers with the highest visible background, we conclude that the upper limit for the remnant PL in the nanoflower spectra is 2\%. 

As mentioned in the main text, the thickness of the nanoflower petal is estimated to be between 2 and \SI{30}{nm}. Therefore we compare the response of the nanoflowers to that of few-layer WS$_2$ in Fig.\ref{fig_backgroundPL}e. The spectra of a trilayer (in blue) and five layers of WS$_2$ (in blue-green) (exfoliated on a Si substrate) exhibit PL both from the direct transition and the indirect transition (around 800 - \SI{850}{nm}). The intensity of the PL from the direct transition is an order of magnitude lower than the PL of the WS$_2$ monolayer, but it is still clearly distinguishable from the background. The intensity of the remnant PL of the nanoflower is however another order of magnitude lower than the PL of few-layer WS$_2$. 

\subsection{Characterization of Raman modes}

\subsubsection{Higher order WS$_2$ Raman modes}

\begin{figure*}[htp]
\centering
\includegraphics[width = 0.7\linewidth] {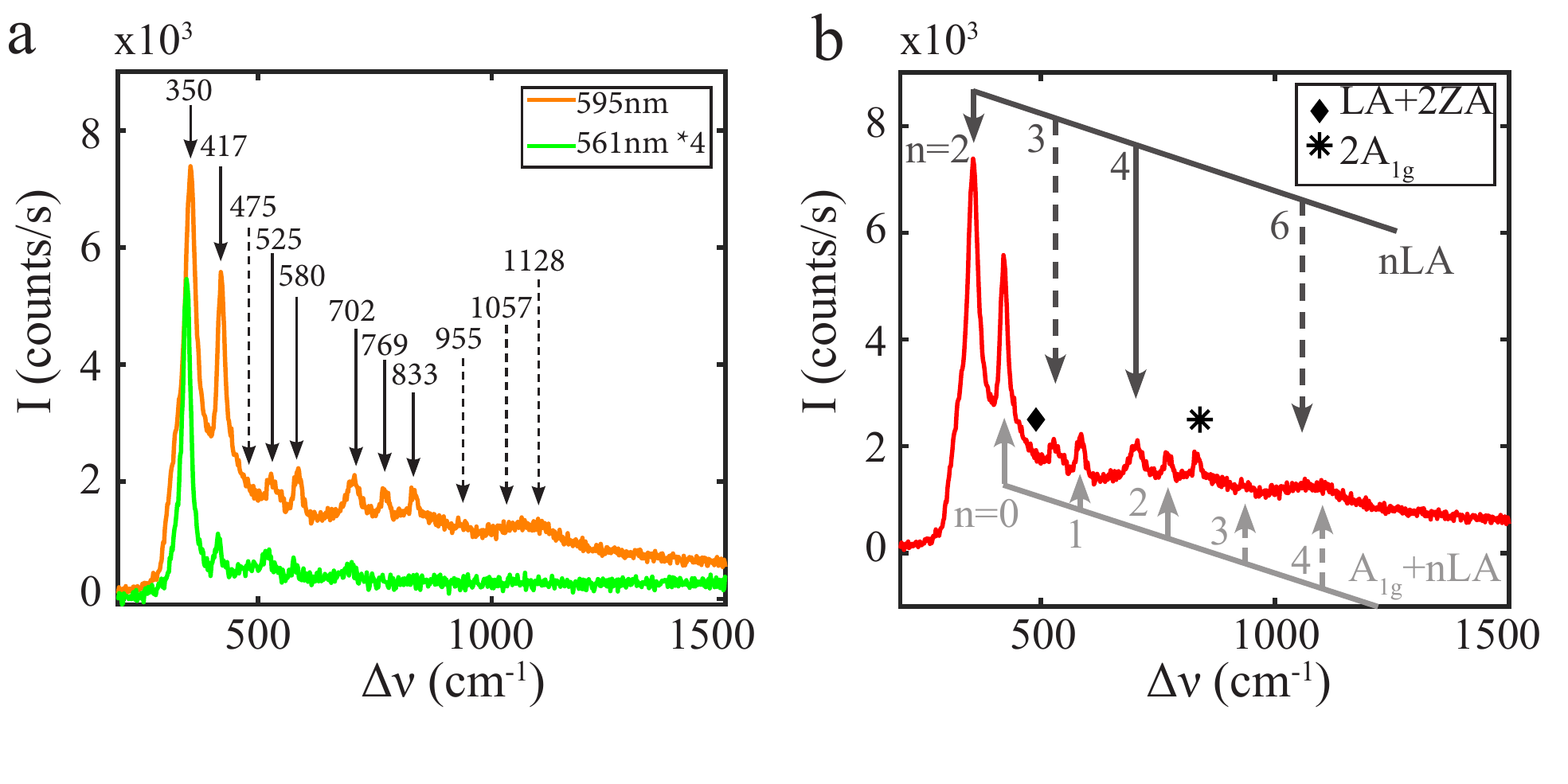}
\caption{\textbf{Characterization of Raman peaks} \\ \textbf{a.} The optical response of the WS$_2$ nanoflowers upon \SI{595}{nm} excitation (in orange) and \SI{561}{nm} excitation (in green, multiplied by 4 for an easy comparison). The spectral response on the two different lasers overlaps well, indicating that the collected light originates from Raman processes. Many more modes are observed using the more resonant \SI{595}{nm} excitation (orange), than using the \SI{561}{nm} excitation (green). The optical response of the WS$_2$ nanoflowers contains 8-10 Raman features. The signature features of WS$_2$, the 2LA(M),$E^1_{2g}$ modes around 350-\SI{355}{cm^{-1}} and the $A_{1g}$ mode at \SI{420}{cm^{-1}}, are clearly observed. \textbf{b.} The other features can be explained as being multiphonon resonances involving the LA(M) phonon (see \cite{Irina_pyramids_2020}). The light grey line under the spectrum depicts the higher order resonances of $A_{1g}$+n*LA(M). The features at \SI{580}{cm^{-1}} and at \SI{769}{cm^{-1}} have been reported before \cite{Molas_RamanWS2_ScientRep_2017, Berkdemir_RamanWS2_ScientRep_2013, Peimyoo_temperatureRamanWS2_NanoRes_2015, Gaur_temperatureRamanWS2_PhysChemC_2015}. As the nanoflower is not located on top of a silicon substrate, we associate the feature around \SI{525}{cm^{-1}} to 3LA(M) rather than to the Si Raman resonance. The dark grey line above the spectrum depicts the higher order resonances of n*LA(M). The feature at \SI{702}{cm^{-1}} has been reported before \cite{Berkdemir_RamanWS2_ScientRep_2013, Peimyoo_temperatureRamanWS2_NanoRes_2015}, as have the features at \SI{475}{cm^{-1}} and \SI{833}{cm^{-1}} \cite{Molas_RamanWS2_ScientRep_2017, McDonnell_resonantRamanWS2_NanoLetters_2018}. The features around \SI{955}{cm^{-1}}, \SI{1057}{cm^{-1}} and \SI{1128}{cm^{-1}} (marked with dotted arrows) cannot be distinguished from the background very well, but have been reported in the spectra of WS$_2$ pyramids \cite{Irina_pyramids_2020}.}
\label{fig_determine}
\end{figure*}

The spectral positions of the sharp features in the spectra of the WS$_2$ nanoflower, taken with a different excitation wavelengths in Fig.\ref{fig_backgroundPL}, do not overlap in wavelength. These features are located at the same relative frequency distance to the excitation laser, as depicted in Fig.\ref{fig_determine}a, indicating that the collected light originates from Raman processes. The positions of the Raman features are indicated with arrows. Commonly, only three Raman modes are measured on both horizontal TMDs layers or nanostructures. Recently, we have reported the attribution of higher frequency Raman modes in spectra of CVD grown WS$_2$ pyramids to multiphonon resonances involving the LA(M) phonon \cite{Irina_pyramids_2020}, adopting the methodology for high frequency Raman features in MoS$_2$ \cite{Golasa_multiphononMoS2_APL_2014}. The light grey line in Fig.\ref{fig_determine}b depicts the higher order resonances of $A_{1g}$+n*LA(M). The features $A_{1g}$+1*LA(M) (at \SI{580}{cm^{-1}}) and $A_{1g}$+2*LA(M) (at \SI{769}{cm^{-1}}) have been reported before \cite{Molas_RamanWS2_ScientRep_2017, Berkdemir_RamanWS2_ScientRep_2013, Peimyoo_temperatureRamanWS2_NanoRes_2015, Gaur_temperatureRamanWS2_PhysChemC_2015}. The dark grey line in Fig.\ref{fig_determine}b depicts the higher order resonances of n*LA(M). The feature 4*LA(M) (at \SI{702}{cm^{-1}}) has been reported before \cite{Berkdemir_RamanWS2_ScientRep_2013, Peimyoo_temperatureRamanWS2_NanoRes_2015}. Although one would also expect a feature at 3*LA(M), this is usually not reported. Most experiments are performed with TMDs on a silicon substrate, and the Si resonance at \SI{520}{cm^{-1}} is around the same position as the mentioned WS$_2$ feature. For this experiment the flowers are positioned on a Si$_3$N$_4$ film far away from the silicon frame, therefore it is safe to assume that the measured feature is not related to Si, but can be attributed to 3LA(M). The features at \SI{475}{cm^{-1}} and \SI{833}{cm^{-1}} have been reported before \cite{Molas_RamanWS2_ScientRep_2017, McDonnell_resonantRamanWS2_NanoLetters_2018}. The features around \SI{955}{cm^{-1}}, \SI{1057}{cm^{-1}} and \SI{1128}{cm^{-1}} (marked with dotted arrows) cannot be distinguished from the background very well, but have been reported in the spectra of WS$_2$ pyramids \cite{Irina_pyramids_2020}. 

\subsubsection{Atomic structure 2H vs 3R}

Naturally occurring WS$_2$ exhibits a hexagonal atomic structure called 2H. However, the scanning transmission electron microscopy (STEM) study reveals that the nanoflowers exhibit a crystallographic polytypism 2H/3R \cite{Sabrya2020}. The 2H and 3R atomic structures can be distinguished comparing the Raman signal of shear- and breathing modes \cite{Lee_MoS2Raman3R_ACSNano_2016, Baren_MoS2Raman3R_2Dmat_2019}. These Raman resonances have frequencies of 10-60 cm$^{-1}$ and therefore lie outside of our experimental spectral region. Differences have been reported in the layer dependent spectral position of the A$_{1g}$ and E$^1_{2g}$ Raman peaks as well as the spectral position of the photoluminescence \cite{Yang_WS2RamanPL3R_Nanotech_2019, Zeng_WS2RamanPL3R_AdvFuncMat_2019}, but the reported differences are too subtle to allow drawing any conclusions based on our measurements. 

\subsubsection{Non-WS$_2$ Raman features}

\begin{figure*}[htp]
\centering
\includegraphics[width = 0.6\linewidth] {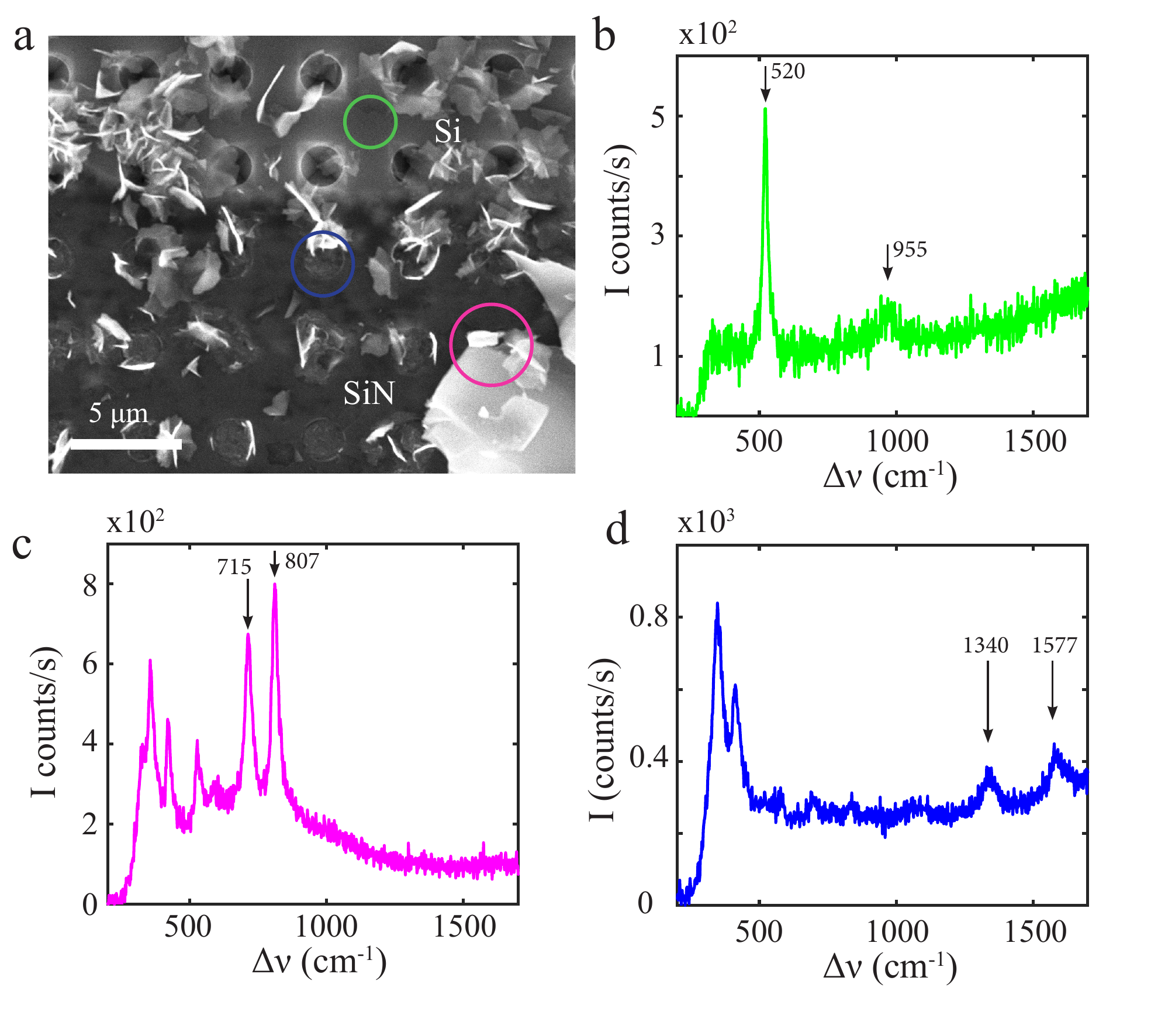}
\caption{\textbf{Non-WS$_2$ Raman response} \\ \textbf{a.} SEM image of the WS$_2$ nanoflowers. The substrate is composed of a silicon frame (upper half of image) with a Si$_3$N$_4$ window in the middle (lower half of image). The array of holes can be distinguished more clearly in the silicon region of the sample, but the holes are also present in the Si$_3$N$_4$ membrane. \textbf{b.} Spectrum on the Si substrate (see green circle in \textbf{a}), where the characteristic \SI{520}{cm^{-1}} and \SI{955}{cm^{-1}} Raman features can be clearly distinguished. These Raman features are not present on the Si$_3$N$_4$ membrane. \textbf{c.} Spectrum of a WO$_3$ particle (see pink circle in \textbf{a}), where the characteristic \SI{715}{cm^{-1}} and \SI{807}{cm^{-1}} Raman features can be clearly distinguished. In most regions of the sample, the only measured Raman features are from WS$_2$ and not WO$_3$. \textbf{d.} Spectrum with the characteristic Raman features of carbon at \SI{1340}{cm^{-1}} and \SI{1577}{cm^{-1}}. We attribute this to the carbon paste that is used to attach the microchip to the sample holder, as the Raman features are only visible at the holes in the Si$_3$N$_4$ membrane (see blue circle in \textbf{a}). }
\label{fig_other_peaks}
\end{figure*}

At some positions, the measured spectra exhibit Raman features from other materials than WS$_2$. As mentioned in the main text, the studied WS$_2$ nanoflowers are fabricated on a Si$_3$N$_4$ membrane with an array of holes. This membrane spans a window in the middle of a silicon. Figure \ref{fig_other_peaks}a depicts an SEM image of a part of the sample, where the WS$_2$ nanoflowers are grown both on the Si frame (upper part of image) and on the Si$_3$N$_4$ membrane (lower part of the image). The holes cross both the Si and the Si$_3$N$_4$. Figure \ref{fig_other_peaks}b depicts a spectrum of the Si substrate (see green circle in Fig.\ref{fig_other_peaks}a), where the characteristic Raman features of Si are clearly present. These Raman features are not present in any other spectra presented in this work, as all the spectra are acquired from nanoflowers on the Si$_3$N$_4$ membrane. 

Preceding the CVD growth procedure of the nanoflowers, WO$_3$ is deposited on the microchip \cite{Sabrya2020}. Signature Raman features of WO$_3$ can be distinguished in the spectrum in Fig.\ref{fig_other_peaks}c, acquired from the large white structure in the right corner of Fig.\ref{fig_other_peaks}a (see pink circle). We conclude that this white structure is a WO$_3$ crystal that has not reacted with sulfur. We do not measure any Raman signatures of WO$_3$ in other regions of the sample. 

Next to Si and WO$_3$, signature Raman features of C can be distinguished in Fig.\ref{fig_other_peaks}d. We attribute this to the carbon paste that is used to attach the microchip to the sample holder. The carbon Raman features are only visible at the holes in the Si$_3$N$_4$ membrane (see blue circle in Fig.\ref{fig_other_peaks}a).  

\subsection{Polarization-resolved Raman response}

\begin{figure*}[htp]
\centering
\includegraphics[width = 0.9\linewidth] {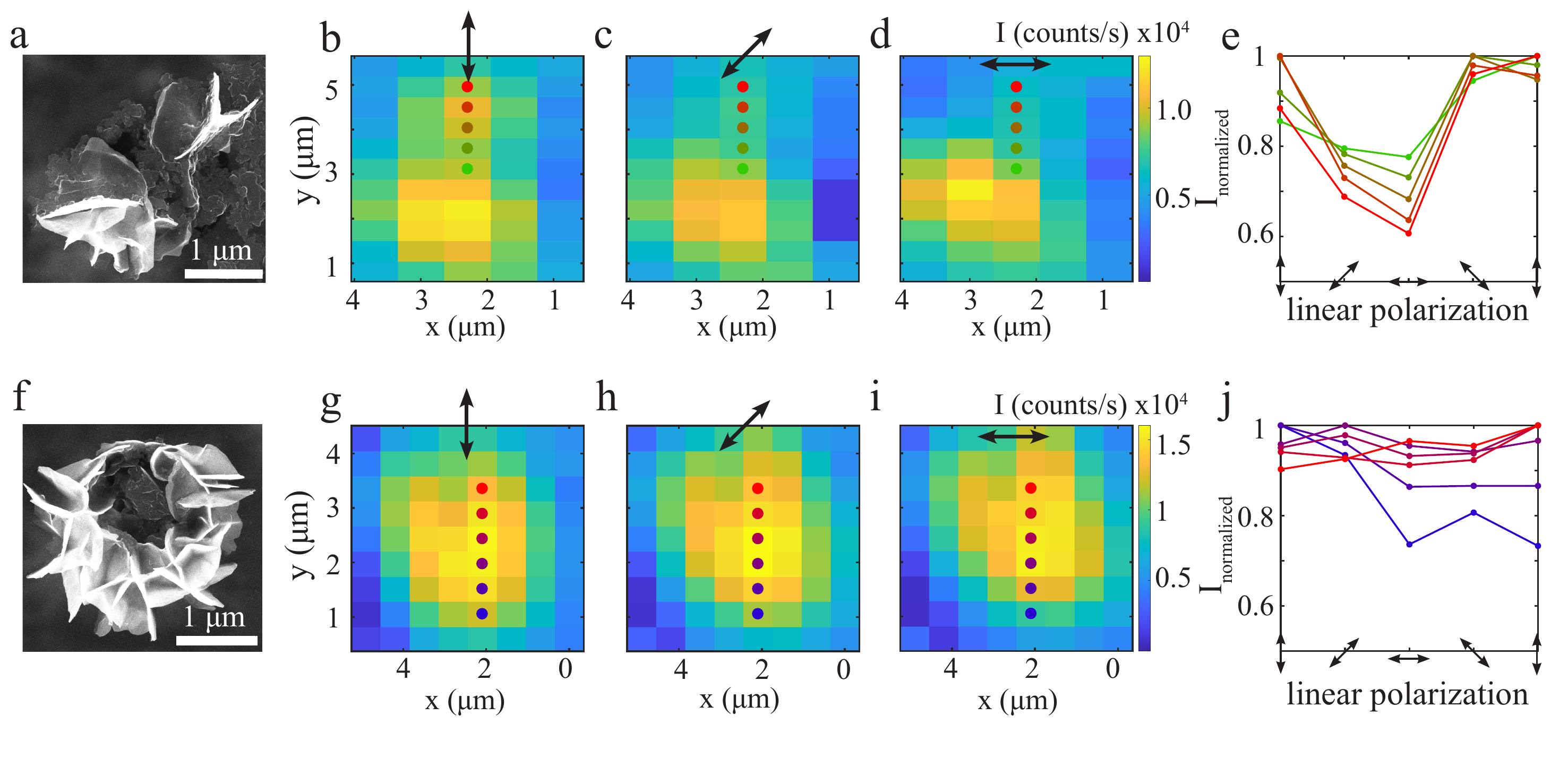}
\caption{\textbf{Excitation polarization} \\
\textbf{a.} SEM image of a WS$_2$ flower-like structure (pink circle in Fig.1a in the main text). The right upper corner of this flower-like structure contains mainly petals oriented in the y-z plane. \textbf{b-d,g-h} Map of the intensity of the first Raman feature of the flower-like structure upon \textbf{b,g.} vertically polarized, \textbf{c,h.} diagonally polarized and \textbf{d,i.} horizontally polarized excitation. \textbf{e.} Raman intensity of the flower in \textbf{a} (used pixels are marked with stars in \textbf{b-d}) as a function of excitation polarization angle. Note that the intensity increases drastically when the polarization direction is parallel to the WS$_2$ flower petals. \textbf{f.} SEM image of a WS$_2$ flower (red circle in Fig.1a in the main text), with petals oriented in all directions. \textbf{j.} Raman intensity of the flower in \textbf{f} (used pixels are marked with stars in \textbf{g-i}) as a function of excitation polarization angle. No polarization dependence can be observed in the Raman intensity of this flower. }
\label{fig_polarization_independence}
\end{figure*}

As mentioned in the main text, we study the interaction of the WS$_2$ nanoflower Raman response with linearly polarized light. \mbox{Figures \ref{fig_polarization_independence}a,f} depict SEM images of flower-like WS$_2$ structures (indicated in pink and red respectively in Fig.1a in the main text). The right upper corner of the flower-like structure in Fig.\ref{fig_polarization_independence}a contains mainly petals oriented in the y-z plane (see coordinate system in Fig.1b in the main text), the petals of the flower in Fig.\ref{fig_polarization_independence}f are oriented in all directions. \mbox{Figures \ref{fig_polarization_independence}b-d and g-i} depict the intensity of the first Raman feature upon vertical polarization excitation, excitation polarization at \SI{45}{degrees} and horizontal polarization excitation. For the upper part of the flower in Fig.\ref{fig_polarization_independence}a, the Raman intensity is highest when the excitation polarization direction is parallel to the orientation of the petals, namely vertically polarized. The Raman intensity of the lower part of this flower, and of the flower in Fig.\ref{fig_polarization_independence}f, does not depend on the excitation polarization direction. Fig.\ref{fig_polarization_independence}e,j depicts the normalized Raman intensity of different parts of the nanoflower (positions are indicated in Fig.\ref{fig_polarization_independence}b-d) as a function of polarization angle (depicted by the arrows). In Fig.\ref{fig_polarization_independence}e, the Raman intensity upon horizontal excitation is 60 - 80 \% of the Raman intensity upon vertical polarization. No polarization dependence can be distinguished in Fig.\ref{fig_polarization_independence}j. 

\begin{figure*}[htp]
\centering
\includegraphics[width = 0.9\linewidth] {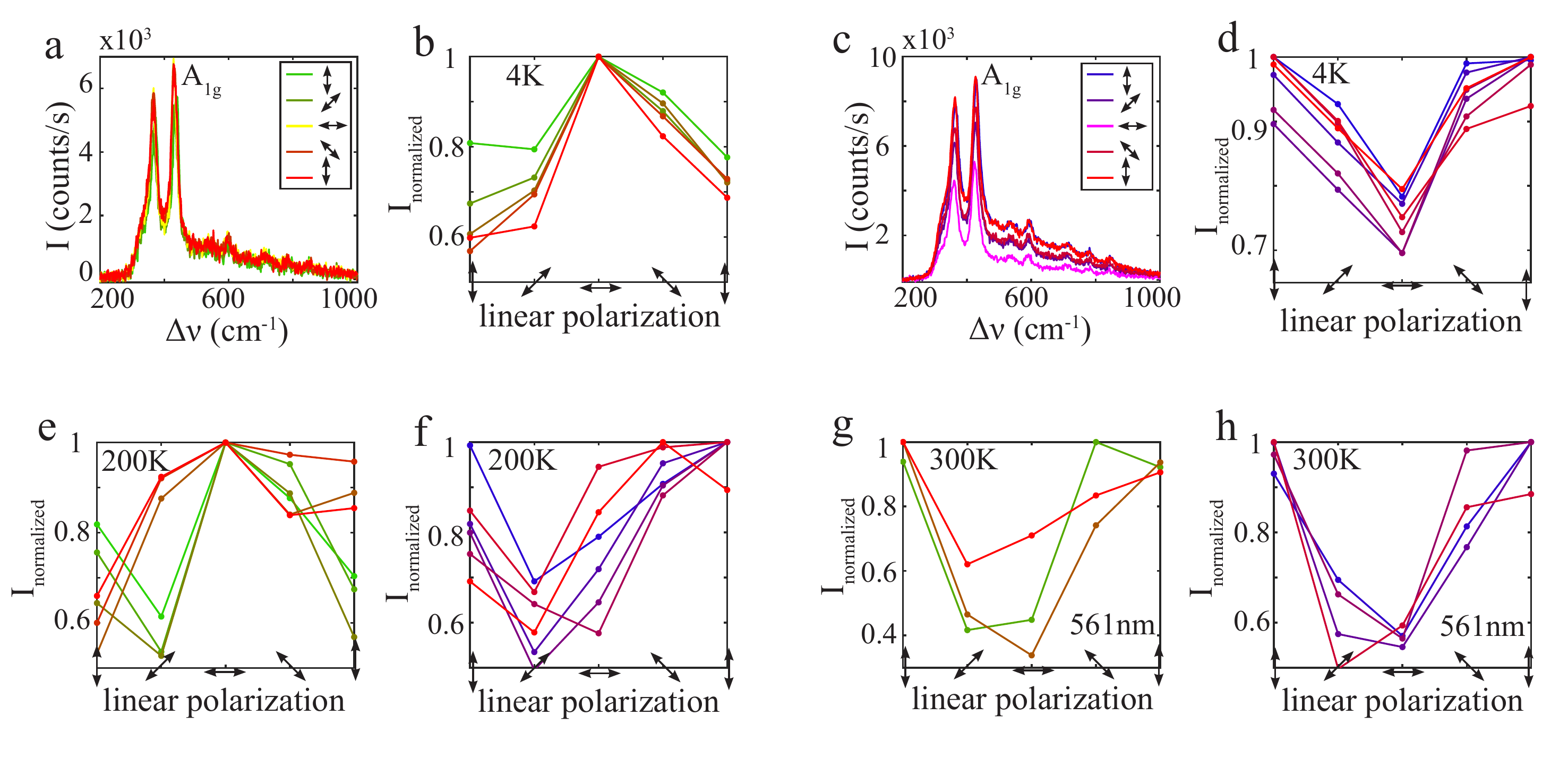}
\caption{\textbf{Excitation polarization} \\
\textbf{a,c.} Polarization-dependent spectra of the WS$_2$ nanoflowers presented in Fig.2a,f in the main text. The intensity of the different Raman features has the same polarization dependence. \textbf{b,d.} Raman intensity of the A$_{1g}$ mode of the flower in Fig.2a,f in the main text as a function of excitation polarization angle. As was the case for the Raman intensity of the first WS$_2$ mode E$_{2g}$,2LA(M), the intensity of the A$_{1g}$ mode increases drastically when the polarization direction is parallel to the WS$_2$ flower petals, e.g., for a horizontal polarization in \textbf{b} (compare Fig.2e in the main text) and a vertical polarization in \textbf{d} (compare Fig.2j in the main text). \textbf{e-h.} Intensity of the first Raman mode as a function of polarization angle of \textbf{e,f.} the flowers in Fig.2a,f in the main text at a temperature of \SI{200}{K} and upon \SI{595}{nm} excitation, and \textbf{g,h} of the flower in Fig.\ref{fig_polarization_independence}a and \textbf{h.} in Fig.2f in the main text, at room temperature upon \SI{561}{nm} excitation. As was the case for the Raman intensity at \SI{4}{K} upon \SI{595}{nm} presented in the main text, the Raman intensity increases when the polarization direction is parallel to the flower petals. Although the noise on the current data is higher, the contrast between parallel and perpendicular polarization is \textbf{e,f} the same for both temperatures upon \SI{595}{nm} excitation: around 0.60 for the flower in Fig.2a and 0.70 for the flower in Fig.2f, and the contrast is \textbf{g,h} slightly larger upon \SI{561}{nm} excitation: around 0.40 in \textbf{g.} and 0.60 in \textbf{h}.}
\label{fig_polarization_modes}
\end{figure*}

Figure 2 in the main text displays the polarization dependence of the intensity of the first Raman feature, the combination of the E$_{2g}$,2LA(M) modes. Figure \ref{fig_polarization_modes}a,c depicts polarization dependent spectra of the  nanoflowers presented in Fig.2 in the main text. The polarization response of the intensity of the first two Raman features is highly similar. This becomes apparent when comparing the polarization-dependent normalized intensity of the second Raman feature, the A$_{1g}$ mode, in Fig.\ref{fig_polarization_modes}b,d to the polarization-dependent response of the first Raman feature in Fig.2e,j in the main text. As the flower in Fig.2a contains mainly petals oriented in the x-z plane, the intensity of both the first and the second Raman feature is highest upon horizontal excitation polarization (Fig.\ref{fig_polarization_modes}b). As the flower in Fig.2f contains mainly petals oriented in the y-z plane, the intensity of two first two Raman modes is highest upon vertical excitation polarization (Fig.\ref{fig_polarization_modes}d). 

Where Fig.2 in the main text displayed the polarization dependence of the Raman intensity at \SI{4}{K}, Fig.\ref{fig_polarization_modes}e,f depicts the polarization dependence of the Raman intensity at \SI{200}{K}. Comparing Fig.\ref{fig_polarization_modes}e,f with Fig.2e,j, it becomes apparent that the Raman intensity at both temperatures increases when the polarization direction is parallel to the flower petals. Although the noise on the data at \SI{200}{K} is higher, the contrast between parallel and perpendicular polarization is the same for both temperatures, namely 0.60 for the flower in Fig.2a and 0.70 for the flower in Fig.2f. We conclude that the polarization-dependence of the Raman intensity does not depend on temperature. 

So far, all mentioned polarization dependences have been using a \SI{595}{nm} excitation. Fig.\ref{fig_polarization_modes}g,h depict the polarization-dependent Raman intensity of the first Raman feature at room temperature upon a \SI{561}{nm} excitation, for the flower in Fig.\ref{fig_polarization_independence}a and in Fig.2j in the main text respectively. As the petals in both cases are mainly oriented in the y-z plane, the Raman intensity is lowest upon horizontal excitation polarization. Although the noise on the data for a \SI{561}{nm} excitation are higher, the contrast between parallel and perpendicular polarization is slightly larger than for a \SI{595}{nm} excitation. 

\subsection{Helicity of Raman features}

\begin{figure*}[htp]
\centering
\includegraphics[width = 0.8\linewidth] {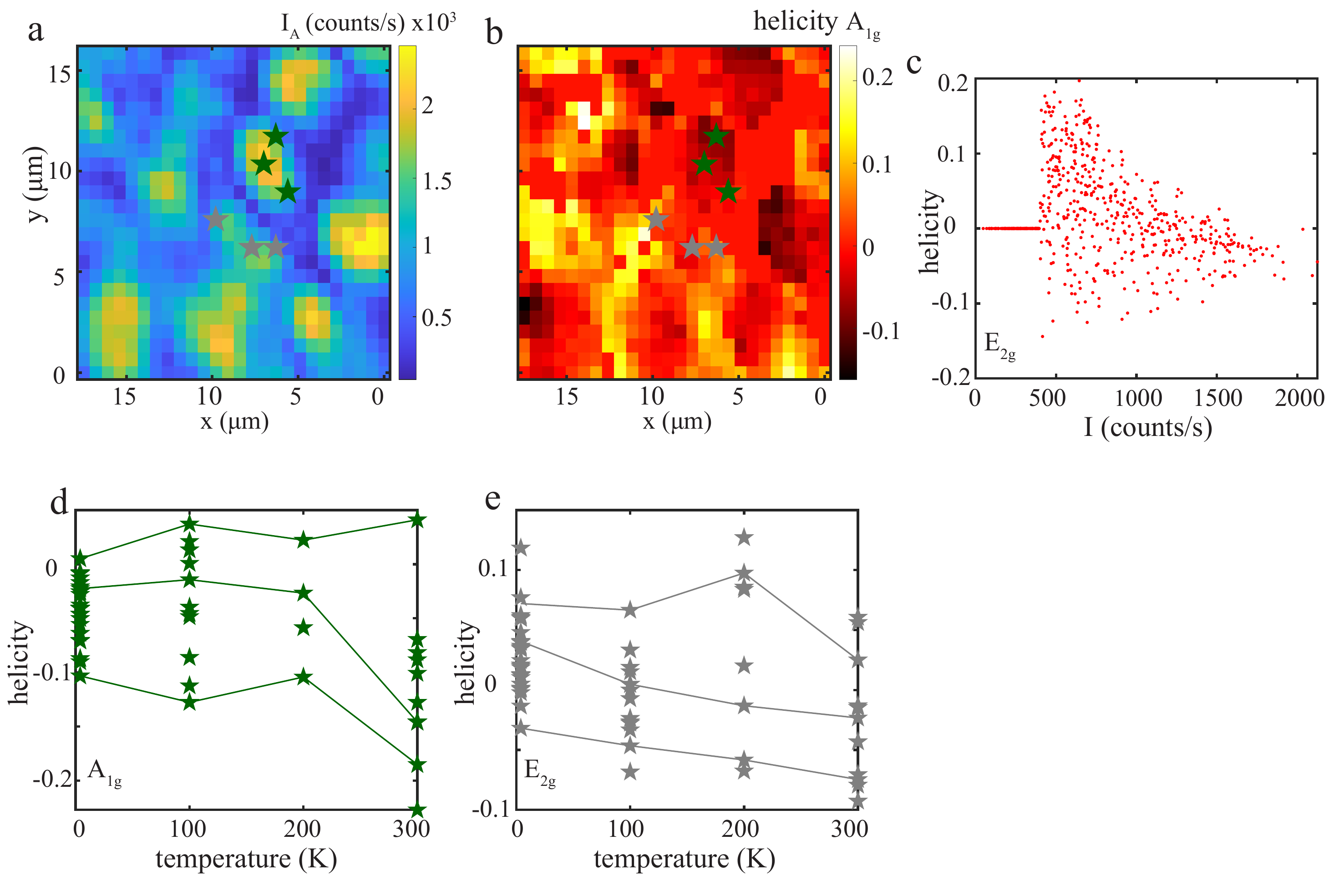}
\caption{\textbf{Raman helicity of A$_{1g}$ mode} \\
\textbf{a.} Map of the intensity of the second Raman feature, the A$_{1g}$, of the nanoflower spectra presented in Fig.3b,c in the main text. \textbf{b.} Map of the same region of the helicity of the A$_{1g}$ mode. The helicity of this Raman feature has a similar value as the first Raman feature presented in the main text, being negative around the WS$_2$ nanoflowers and positive in regions next to the larger nanoflowers (see Fig.3e in the main text). \textbf{c.} Distribution of the Raman helicity (of the first Raman feature, presented in the main text), as a function of Raman intensity. At low intensity, the helicity can take a broad range of values between -0.1 and +0.2. At high intensity, the helicity goes to a value of around -0.05. \textbf{d,e.} Temperature-dependent helicity \textbf{d.} of the A$_{1g}$ mode and \textbf{e.} of the first Raman feature of the WS$_2$ nanoflower marked in \textbf{d.} green and \textbf{e.} grey in \textbf{a,b.} (taking into account all pixels associated with this flower). The lines present the temperature dependence of three specific places on the flowers, denoted with stars in \textbf{a,b}. Like in Fig.3f in the main text, the helicity decreases slightly at room-temperature.}
\label{fig_helicity_other}
\end{figure*}

Figure 3d-f in the main text displayed the intensity and helicity of the first WS$_2$ Raman feature, the combination of the E$_{2g}$,2LA(M) modes. Figure \ref{fig_helicity_other}a presents a map of the nanoflower intensity of the second Raman feature, the A$_{1g}$ mode. Figure \ref{fig_helicity_other}b presents a map of the experimentally determined helicity of the A$_{1g}$ mode. When comparing Fig.\ref{fig_helicity_other}b with Fig.3e in the main text, note that the helicity of the two Raman features has very similar position-dependent values. Comparing the position of the (bright) nanoflowers on the intensity map with the helicity map, it becomes apparent that the Raman helicity of the nanoflowers is slightly negative. Figure \ref{fig_helicity_other}c depicts the helicity of the first Raman feature as a function of intensity. At low intensity, the helicity values are spread in a range from -0.10 and +0.20. At high intensity, the spread in the helicity becomes smaller and converges to a value around -0.05. 

Figure \ref{fig_helicity_other}d depicts the temperature dependence of the helicity of the A$_{1g}$ mode for the flower marked in green in Fig.1a. At all temperatures, the intensity is depicted of the first Raman feature of all spectra associated to this flower. The lines present the temperature dependence of three specific places on the flower, marked by green stars in Fig.\ref{fig_helicity_other}a,b. Figure \ref{fig_helicity_other}e depicts the temperature dependence of the helicity of the first Raman mode of another flower. The lines present the temperature dependence of three places on the flower, marked by grey stars in Fig.\ref{fig_helicity_other}a,b. As for the flower and the Raman feature in the main text, the helicity of these flowers slightly decreases from \SI{4}{K} to room temperature. 

\begin{figure*}[htp]
\centering
\includegraphics[width = 0.8\linewidth] {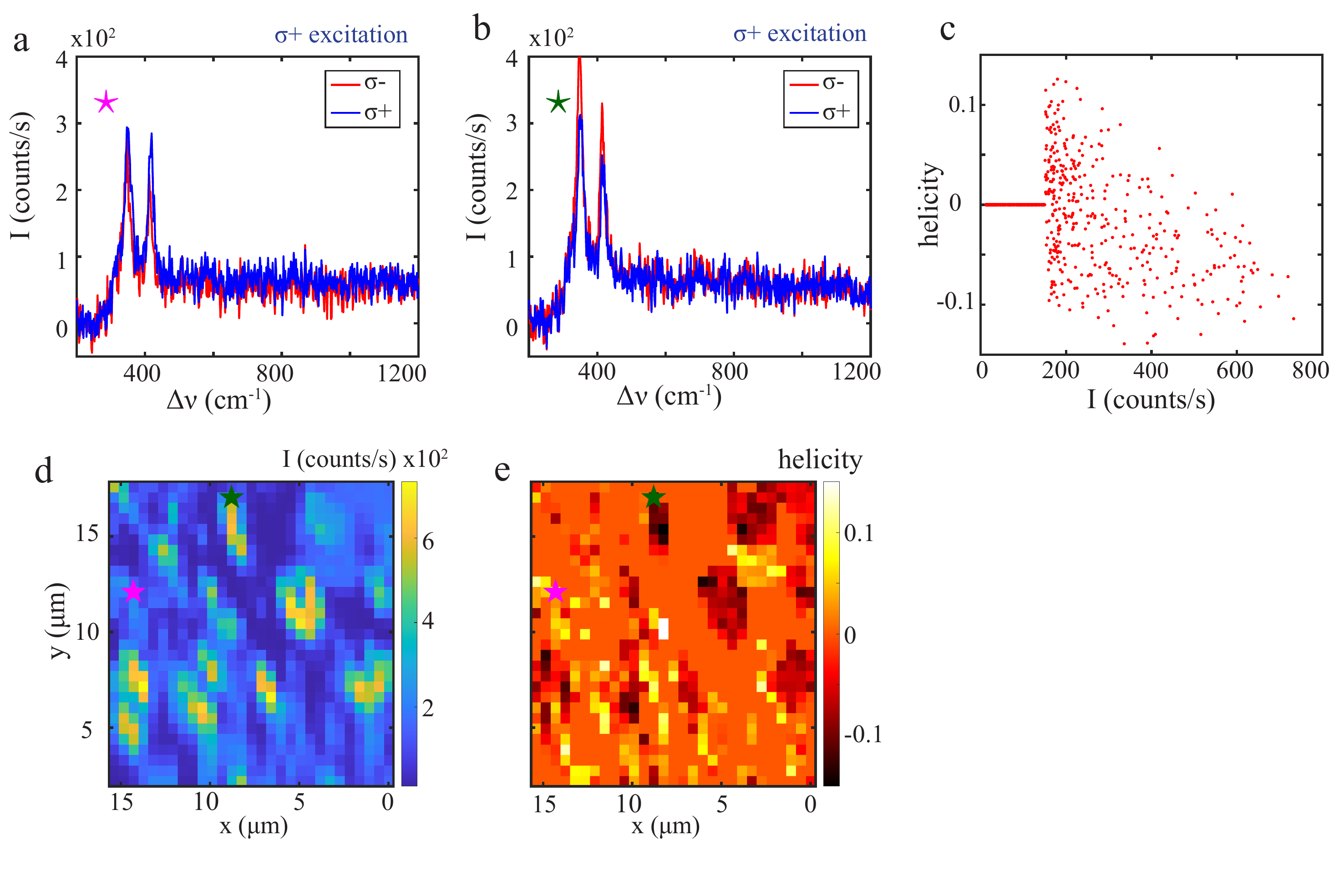}
\caption{\textbf{Raman helicity at room temperature} \\ 
\textbf{a,b} Helicity-resolved nanoflower spectra, where the flowers are excited with $\sigma_+$ light and the helicity is determined from the difference in $\sigma_+$ and $\sigma_-$ emission. The spectra are taken at room temperature at the same positions as in Fig.3 in the main text. In \textbf{a}, the spectrum with the same polarization as the excitation light (blue) has a higher intensity (helicity is conserved). In \textbf{b}, the spectrum with opposite polarization to the excitation light (red) has a higher intensity (helicity is reversed). \textbf{c.} Distribution of the helicity of the first Raman feature as a function of Raman intensity. At low intensity, the helicity can take a broad range of values between -0.15 and +0.1. At high intensity, the helicity goes to a value of around -0.1. \textbf{d.} Map of the intensity of the first Raman feature, taken at room temperature. \textbf{e.} Map of the same region of the Raman helicity. As in Fig.3, the helicity of the Raman features around the WS$_2$ nanoflowers is negative (green star), whereas the Raman helicity is positive in regions next to the larger nanoflowers (pink star). }
\label{fig_helicity_300K}
\end{figure*}

For comparison, we determine the position-dependent helicity of the Raman features at different temperatures. Figure \ref{fig_helicity_300K}a,b depict helicity-resolved nanoflower spectra, taken at room temperature. Here, the flowers are excited with $\sigma_+$ light and the helicity of the Raman features is determined from the difference in $\sigma_+$ and $\sigma_-$ emission (see Fig.3a in the main text). In Fig.\ref{fig_helicity_300K}a, the blue spectrum with the same polarization as the excitation light, has a higher intensity (helicity is conserved) than the red spectrum with the opposite polarization (helicity is reversed). In Fig.\ref{fig_helicity_300K}b, the helicity-reversed spectrum (in red) has a higher intensity than the helicity-conserved spectrum (in blue). 

Figure \ref{fig_helicity_300K}d,e present a map of the nanoflower intensity and helicity of the first Raman feature, taken at room temperature. The stars indicate the position of the spectra in Fig.\ref{fig_helicity_300K}a,b (compare Fig.3d,e in the main text). As was the case at \SI{4}{K} (see Fig.3 in the main text), at room temperature the Raman helicity is also negative at the position of the WS$_2$ nanoflowers, and positive on locations in between the flowers (compare position of green and pink star in Fig.\ref{fig_helicity_300K}d,e). As depicted in Fig.3f in the main text, the Raman helicity is on average more negative at room temperature than at cryogenic temperatures. Figure \ref{fig_helicity_300K}c depicts the helicity of the first Raman feature as a function of intensity. As was the case at \SI{4}{K}, the spread in the helicity values is large for low intensities, but the spread becomes smaller at high intensity. Comparing Fig.\ref{fig_helicity_other}c and Fig.\ref{fig_helicity_300K}c, it becomes apparent that the helicity at room temperature has lower values: the maximum helicity is only 0.12 instead of 0.20, and the value at high intensity is -0.10 instead of -0.05. 

\subsection{Raman polarizability tensors}

\subsubsection{Tensors and Jones calculus}

As mentioned in the main text, the interaction of TMDs materials with polarized light can be described by Raman polarizability tensors. The Jones vectors for circularly polarized light are \cite{Zhao_helicityMoS2_ACSNano_2020}: 

\begin{equation}
    \begin{split}
        \sigma_+ = 1/\sqrt{2} \begin{bmatrix} 1\\i\\0 \end{bmatrix} \textrm{and} \;
        \sigma_- = 1/\sqrt{2} \begin{bmatrix} 1\\-i\\0 \end{bmatrix}.
    \end{split}
\end{equation}

The first two Raman features of WS$_2$ are the combination of the E$_{2g}$ and 2LA(M) mode, and the A$_{1g}$ mode. We start with the Raman tensor of the second feature, the A$_{1g}$ mode \cite{Chen_helicityRamanTMD_NanoLett_2015}:

\begin{equation}
R_A = \begin{bmatrix}
        a & 0 & 0 \\
        0 & a & 0 \\
        0 & 0 & b
    \end{bmatrix} .
\end{equation}

If the incident and outgoing light have the same polarization handedness, the calculation yields a non-zero matrix element \cite{Chen_helicityRamanTMD_NanoLett_2015}: \mbox{$\sigma_{+}^\dagger R_A \sigma_{+} = a $.} If the incident and outgoing light have different polarization handedness, the matrix element is zero \cite{Chen_helicityRamanTMD_NanoLett_2015}: $\sigma_{+}^\dagger R_A \sigma_{-}  = 0$.

Calculating the polarization response of the E$_{2g}$ mode is less straighforward, as the Raman tensors depend on the resonance conditions of the excitation light with the excitonic transition \cite{Zhao_helicityMoS2_ACSNano_2020}. If the excitation is out-of-resonance, the Raman tensor is \cite{Chen_helicityRamanTMD_NanoLett_2015}:

\begin{equation}
R_E = \begin{bmatrix}
        0 & d & 0 \\
        d & 0 & 0 \\
        0 & 0 & 0
    \end{bmatrix} .
\end{equation}

In this case, if the incident and outgoing light have the same polarization handednes, the matrix element is zero \cite{Chen_helicityRamanTMD_NanoLett_2015}: $\sigma_{+}^\dagger R_E \sigma_{+} = 0 $. If the incident and outgoing light have different polarization handedness, the matrix element is non-zero: $\sigma_{+}^\dagger R_E \sigma_{-} = d $. 

In the main text, we calculate the helicity of the measured Raman features: \mbox{$H = \frac{I_{conserved} - I_{reversed}}{I_{conserved} + I_{reversed}}$}, where $I_{conserved}$ had a $\sigma_+$ and $I_{reversed}$ a $\sigma_-$ polarization. Note that the helicity is calculated based on intensities, therefore the matrix elements need to be squared. We calculate the helicity as:

\begin{equation}
    H = \frac{I_{\sigma+\sigma+}-I_{\sigma+\sigma-}}{I_{\sigma+\sigma+}+I_{\sigma+\sigma-}} .
\end{equation}

The helicity of the A$_{1g}$ mode is \mbox{$H = \frac{a^2-0}{a^2+0} = +1$}, the helicity is conserved. The helicity of the E$_{2g}$ mode is \mbox{$H = \frac{-d^2}{d^2} = -1$}, the helicity is reversed. In summary: when the excitation light is out of resonance with the excitonic transition of a TMDs material, the first two Raman features respond to circular polarization in an opposite way: the helicity is reversed for the first feature, and conserved for the second. 

\subsubsection{Resonant excitation}

The measured helicity-resolved Raman of the WS$_2$ nanoflowers exhibits a different type of behaviour. We experimentally observe that the two first Raman features exhibit the same response to circularly polarized light, either both being helicity conserved or helicity reversed. Part of this difference between theory and experiment can be explained by the fact that the incident light on the nanoflowers is in resonance with the excitonic transition. 

In this case, the Raman tensor for the A$_{1g}$ mode remains the same, but the polarization response of the E$_{2g}$ mode has two contributions \cite{Zhao_helicityMoS2_ACSNano_2020, Drapcho_helicityTMD_PRB_2017}. The interaction between electrons, photons and excitons is governed by the so called deformation potential (DP) and Frohlich interaction (FI) \cite{Zhao_helicityMoS2_ACSNano_2020}, leading to the following Raman tensors: 

\begin{equation}
\label{equat_tensors_resonant}
\begin{split}
R_{LO} = 
    \begin{bmatrix} 
        a_F & a_{DP} & 0 \\ 
        a_{DP} & a_F & 0 \\ 
        0 & 0 & a_F 
        \end{bmatrix} ,
R_{TO} = 
    \begin{bmatrix}
        a_{DP} & 0 & 0 \\
        0 & -a_{DP} & 0 \\
        0 & 0 & 0
    \end{bmatrix} .
\end{split}
\end{equation}

\begin{figure*}[htp]
\centering
\includegraphics[width = 0.6\linewidth] {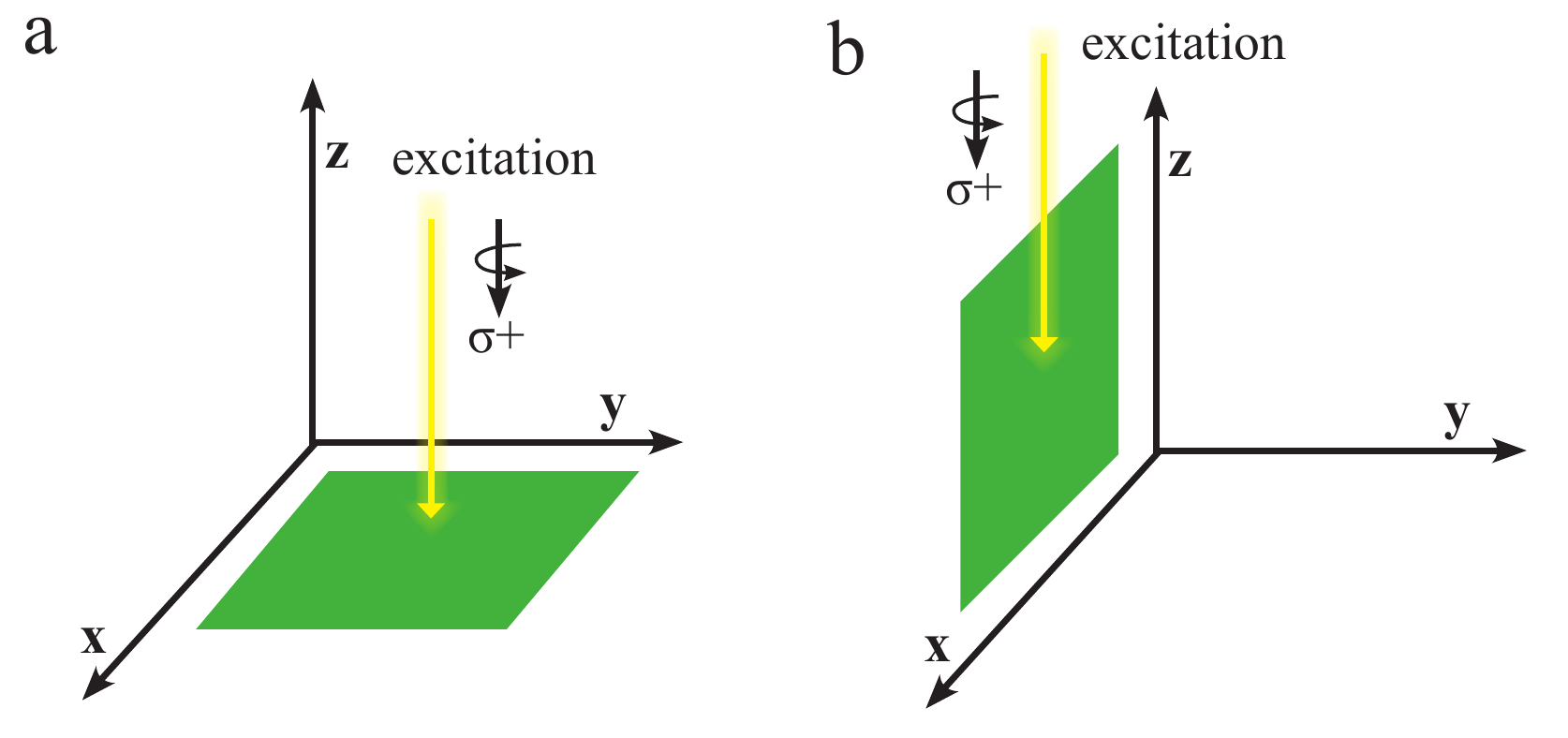}
\caption{\textbf{Coordinate systems} \\ 
\textbf{a.} Coordination system with a flat WS$_2$ layer in the x-y plane, and circularly polarized excitation propagating along z. \textbf{b.} Exciting a wall-like WS$_2$ flower petal is like performing a base transform. Here the new base vectors are: $\hat{x}' = \hat{x}$, $\hat{y}' = -\hat{z}$ and $\hat{z}' = \hat{y}$. }
\label{fig_base_transformation}
\end{figure*}

If the incident and outgoing light have the same polarization handedness, \mbox{$\sigma_{+}^\dagger R_{LO} \sigma_{+} = a_F $} and \mbox{$\sigma_{+}^\dagger R_{TO} \sigma_{+} = 0 $}. If the incident and outgoing light have different polarization handedness, \mbox{$\sigma_{+}^\dagger R_{LO} \sigma_{-} = -ia_{DP} $} and \mbox{$\sigma_{+}^\dagger R_{TO} \sigma_{-} = a_{DP}$}. Therefore, independent on the polarization handedness, the interaction will always contain a non-zero matrix element. The helicity of the E$_{2g}$ mode is \mbox{$H = \frac{a_F^2-2a_{DP}^2}{a_F^2+2a_{DP}^2}$}. Depending on the contribution of the DP and FI interactions, the helicity of the E$_{2g}$ mode can be either conserved or reversed \cite{Zhao_helicityMoS2_ACSNano_2020}. 

In summary: when the excitation light is in resonance with the excitonic transition of a TMDs material, both the E$_{2g}$ and the A$_{1g}$ mode can be helicity conserved (H $>$ 0) \cite{Zhao_helicityMoS2_ACSNano_2020, Drapcho_helicityTMD_PRB_2017}.

\subsubsection{Base transformation}

Still, the theory above does not adequately describe the measured helicity-resolved Raman of the WS$_2$ nanoflowers exhibits. Although the resonance condition explains why both Raman features exhibit the same helicity response, it does not explain the observed helicities between -0.20 and +0.20 for the A$_{1g}$ mode, where a helicity of +1.0 would be expected. As mentioned in the main text, the Raman polarizablity tensors are defined with respect to the crystal axes of flat TMDs layers, e.g., a frame of reference with the excitation light perpendicular to it. As the petals of the WS$_2$ flowers exhibit a variety of orientations with respect to the incident light, the Raman tensor needs to be defined in a different frame of reference (see \cite{Jin_MoSe2polarization_2020, Ding_RamanTensorsMoS2_optlett_2020, Hulman_MoS2polarizationVertical_PhysChemC_2019}). Figure \ref{fig_base_transformation}a presents schematically a WS$_2$ flake in the horizontal x-y plane, excited by circularly polarized light propagating along z. Figure \ref{fig_base_transformation}b presents a wall-like WS$_2$ petal in the x-z plane. The WS$_2$ Raman tensor needs to be defined in this rotated coordinate system, where the base vectors transform to: $\hat{x}' = \hat{x}$, $\hat{y}' = -\hat{z}$ and $\hat{z}' = \hat{y}$. The Raman tensor of the A$_{1g}$ mode will change to: 

\begin{equation}
\label{equat_Amode}
R_A' = \begin{bmatrix}
        a & 0 & 0 \\
        0 & -a & 0 \\
        0 & 0 & -b
    \end{bmatrix} .
\end{equation}

If the incident and outgoing light have the same polarization handedness, the matrix element is zero: $\sigma_{+}^\dagger R_A' \sigma_{+} = 0 $. If the incident and the outgoing light have different polarization handedness, the matrix element is non zero: $\sigma_{+}^\dagger R_A' \sigma_{-} = 2a $. Note that now the A$_{1g}$ mode is helicity reversed instead of helicity conserved: H=-1.0.

Applying the same base transformation to the tensors of the E$_{2g}$ mode yields:

\begin{equation}
\begin{split}
R_{LO}' = 
    \begin{bmatrix} 
        a_F & 0 & -a_{DP} \\
        0 & a_F & 0 \\ 
        -a_{DP} & 0 & a_F
        \end{bmatrix} , \\
R_{TO}' = 
    \begin{bmatrix}
        a_{DP} & 0 & 0 \\
        0 & 0 & 0 \\
        0 & 0 & -a_{DP}
    \end{bmatrix} .
\end{split}
\end{equation}

If the incident and outgoing light have the same polarization handedness, the matrix element are: \mbox{$\sigma_{+}^\dagger R_{LO}' \sigma_{+} = a_F $} and \mbox{$\sigma_{+}^\dagger R_{TO}' \sigma_{+} = 1/2a_{DP} $}. If the incident and outgoing light have different polarization handedness, the matrix elements are: \mbox{$\sigma_{+}^\dagger R_{LO}' \sigma_{-} = 0 $} and \mbox{$\sigma_{+}^\dagger R_{TO}' \sigma_{-} = 1/2a_{DP} $}. Therefore the helicity again depends on the contribution of the DP and FI interactions \cite{Zhao_helicityMoS2_ACSNano_2020}: $H = \frac{a_F^2}{a_F^2+1/2a_{DP}^2}$.

\bibliographystyle{unsrt}
\bibliography{article_flowers}

\end{document}